\documentclass{article}

\usepackage{arxiv}
\usepackage[colorlinks=true,
            linkcolor=red,
            anchorcolor=blue,
            citecolor=blue,
            ]{hyperref}
\usepackage{booktabs}
\usepackage[utf8]{inputenc} 
\usepackage[T1]{fontenc}    
\usepackage{hyperref}       
\usepackage{url}            
\usepackage{booktabs}       
\usepackage{amsfonts}       
\usepackage{nicefrac}       
\usepackage{microtype}      
\usepackage{lipsum}
\usepackage{amsmath}
\usepackage{graphicx}
\usepackage[round]{natbib}
\usepackage[ruled,linesnumbered]{algorithm2e}
\usepackage[toc,page]{appendix}
\usepackage{subfigure}
\usepackage{float}

\title{CatBoost model with synthetic features in application to loan risk assessment of small businesses}

\author{
  \href{https://orcid.org/0000-0003-1589-0109}{\includegraphics[scale=0.06]{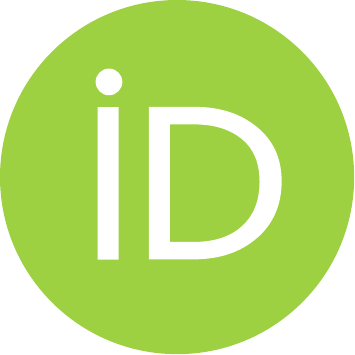}\hspace{1mm}
  Haoxue Wang}\thanks{The corresponding author, the article is contributed equally}\\
  Department of Mathematics\\
  Sichuan University\\
  Chengdu, China \\
  \texttt{wanghaoxue@stu.scu.edu.cn} \\
  \And
  {Liexin Cheng}\\
  Department of Mathematics\\
  Sichuan University\\
  Chengdu, China \\
  \texttt{chenglx@stu.scu.edu.cn} \\
}

\begin{document}
\maketitle

\begin{abstract}
Loan risk for small businesses has long been a complex problem worthy of exploring. Predicting the loan risk can benefit entrepreneurship by developing more jobs for the society. CatBoost (Categorical Boosting) is a powerful machine learning algorithm suitable for dataset with many categorical variables like the dataset for forecasting loan risk. In this paper, we identify the important risk factors that contribute to loan status classification problem. Then we compare the performance between boosting-type algorithms(especially CatBoost) with other traditional yet popular ones. The dataset we adopt in the research comes from the U.S. Small Business Administration (SBA) and holds an enormous sample size (899,164 observations and 27 features). In order to make the best use of the important features in the dataset, we propose a technique named  "synthetic generation" to develop more combined features based on arithmetic operation, which ends up improving the accuracy and AUC of the original CatBoost model. We obtain a high accuracy of 95.84\% and well-performed AUC of 98.80\% compared with the existent literature of related research.
\end{abstract}

\keywords{CatBoost \and Synthetic features \and Loan risk \and Machine Learning \and Default rate}

\section{Introduction}
Our research is focused on the loan risk of small businesses that are supposed to be the primary source of job creation in the United States. Stimulating the establishment and growth of small businesses contributes to creating job opportunities and decreasing unemployment. There are organizations like U.S. Small Business Administration (SBA) that encourage banks to grant loans to small businesses by sharing part of the risk. If a loan default occurs, these intermediary organizations will pay the portion they guaranteed. Since the interest of loaners and obligors are bonded together, it is crucial for banks and corresponding organizations to evaluate the credit of small businesses. The more accurately they identify the obligors, the more profits they make. One way to solve the decision making problem is through analyzing corresponding historical data of the small businesses that apply for the loans, which also forms a famous and heated field of study called credit scoring.

To identify a proper method for loan default detection tasks like this, we introduce CatBoost, currently the most advanced gradient boosting algorithm, along with linear models like logistic regression and nonlinear models such as Support Vector Machine (SVM) and neural networks. Through an empirical experiment, we compare the popular classification methods above and discuss the results.

\section{Previous works}
Terminology like bankruptcy prediction, credit rating/scoring, and corporate financial distress forecast actually refer to the same task and they will be considered as “financial credit risk assessment” \citep{2010Bueyuekkarabacak}. Several empirical researches have identified various characteristics of credit risk assessment. For example, it has been demonstrated that household credit expansions have been a critical indicator of banking crises both statistically and economically. Enterprise credit expansions are also correlated with banking crises, but their effect is weaker and less robust \citep{2016Financial}. 

Financial credit risk assessment originated from more than two centuries ago when most assessments were done qualitatively. Quantitative and less subjective techniques were not popular until the 20$-th$ century. There were some classic models such as the seminal univariate analysis work of Beaver \citep{1966Financial} and multiple discriminant analysis works of Altman in the 1960s \citep{1968FINANCIAL}. Their work demonstrated the ability to predict a company’s failure up to five years in advance. Today, the task is approached using statistical models or advanced machine learning methods. These techniques include artificial neural networks (ANNs) \citep{Guoqiang1999Artificial, 1993A}, fuzzy set theory (FST) \citep{2006Bankruptcy}, decision trees \citep{2006Genetic}, rule-based reasoning \citep{2013A}, support vector machines (SVMs) \citep{2006An}, rough set theory (RST) \citep{1999Business}, genetic programming (GP) \citep{2006Genetic}, hybrid learning \citep{2010A}, and ensemble learning among others \citep{2006An}. Works of Anthony Bellotti \& Damiano Brigo \citep{2019ForecastingBellotti} showed that rule-based algorithms, for example, Cubist, boosted trees and random forests, exhibited significantly better results than other approaches when predicting loan recovery rates. Hung \& Chen \citep{2009A} proposed a selective ensemble of three classifiers, i.e. the decision tree, the backpropagation neural network and the support vector machine, and found that the selective ensemble performs better than other types of ensembles for credit risk assessment. Further, Alfaro \citep{2008Alfaro} compared the performance of AdaBoost and neural networks and indicated that the former presented better results in bankruptcy prediction. This is not an individual case. Zieba \& Tomczak \citep{2016EnsembleZieba} explored the extreme gradient boosting framework on a dataset from Polish companies and gained results significantly better than those by all its reference methods that were applied to the problem before. The success of boosting algorithms, both adaptive boosting and gradient boosting, has sufficiently shown the superiority of ensemble learning (typically combined with decision trees) in predicting credit risk. 

Compared with the assessment of loan default risk of individual obligors, the analysis of financial credit risk of a company can be regarded as a similar task, but with more categorical factors to consider. Banks do not gather detailed information about companies’ financial conditions, but require data such as the city name, industrial classification of borrowers, loan amount and loan term.

Here we propose a novel Gradient boosting framework that claims to exceed the performance of other gradient boosting algorithms including XGBoost, LightGBM. The new boosting algorithm has the advantage of avoiding overfitting of prediction and effectively dealing with categorical features, which can take the form of city names and user IDs. As far as we know, there has been rather limited work on credit risk prediction using CatBoost. The works of Jabeur \citep{2021Ben} demonstrated the superiority of CatBoost in small data bankruptcy prediction. Our work takes an improved approach using synthetic features and evaluates the performance under a more comprehensive dataset. Specifically, our contributions are as follows.
\begin{itemize}
  \item [1]
  Compare the performance of CatBoost with other popular algorithms on small business loan risk evaluation. 
  \item [2]
  Evaluate the power of CatBoost when dealing with categorical features and illustrate the improved performance.
  \item [3]
  Improve the performance of CatBoost by integrating the method of synthetic features.
\end{itemize}

\section{Data}
In this section, we will describe the source of dataset that we use for our model validation. Then we will describe and explain the variables in the dataset as well as specific noticeable characteristics. Some of the data are missing and variables are heterogeneous, i.e. including both numerical and categorical features. We will discuss how to approach these problems. 


\subsection{Dataset}
The dataset \textit{national SBA} is from the U.S. Small Business Administration (SBA) and includes historical data from 1987 through 2014 (899,164 observations). As the name suggests, \textit{national SBA} is aimed at small businesses across America. 

\begin{figure}[!ht]
    \centering
    \includegraphics[width = 10.3cm]{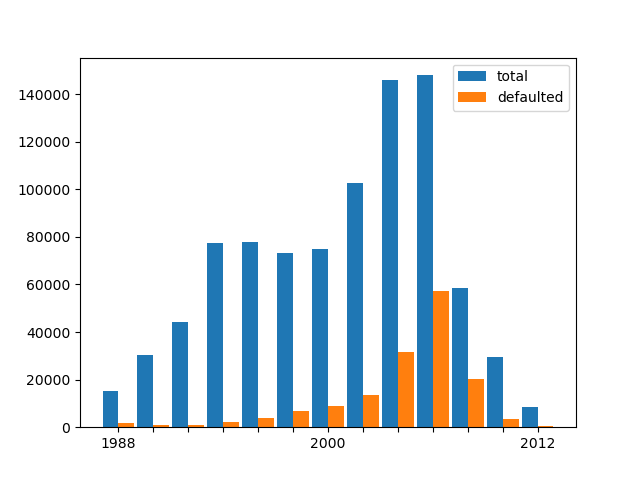}
    \caption{Distribution of total and defaulted loans by year}
    \label{fig:my_label}
\end{figure}
Given the enormous amount of observations and the length of collecting period, we narrow down the scale of dataset for more accessible model performing. First, the borrowing items with missing information in key variables are left out, such as those with missing bank names or city names. Of course there are also data entry errors. For example, in variable \textit{LowDoc}, which is expected to take only two values “Y”, indicating loans with a one-page application, and “N”, indicating loans with more information attached to the application, the dataset contains other symbols like "0" and “S”. These errors account for 0.38\% and should be removed. We also concentrate on observations from 2003 to 2008 since loan businesses are active (see figure \ref{fig:my_label}) and changes in the economic environment are limited. Particularly, the term from December 2007 to June 2009, when a significant recession in global economy happened, should be avoided. After the operations above, the dataset has around 221.5 thousand observations left. The description of 27 variables in the dataset is shown in table \ref{table1} \citep{table1}.
\begin{table}[ht]
  \caption{Description of 27 variables in the dataset}
  \centering
    \begin{tabular}{ccc}\midrule
    Variable name     & Data type & Description of variable                              \\ \midrule
    LoanNr\_ChkDgt    & Text      & Identifier – Primary key                           \\
    Name              & Text      & Borrower name                                        \\
    City              & Text      & Borrower city                                        \\
    State             & Text      & Borrower state                                       \\
    Zip               & Text      & Borrower zip code                                    \\
    Bank              & Text      & Bank name                                            \\
    BankState         & Text      & Bank state                                           \\
    NAICS             & Text      & North American industry classification system code \\
    ApprovalDate      & Date/Time & Date SBA commitment issued                           \\
    ApprovalFY        & Text      & Fiscal year of commitment                            \\
    Term              & Number    & Loan term in months                                  \\
    NoEmp             & Number    & Number of business employees                         \\
    NewExist          & Text      & 1=Existing business, 2=New business                  \\
    CreateJob         & Number    & Number of jobs created                               \\
    RetainedJob       & Number    & Number of jobs retained                              \\
    FranchiseCode     & Text      & Franchise code, (00000 or 00001)=No franchise        \\
    UrbanRural        & Text      & 1=Urban, 2=rural, 0=undefifined                      \\
    RevLineCr         & Text      & Revolving line of credit: Y=Yes, N=No                \\
    LowDoc            & Text      & LowDoc Loan Program: Y=Yes, N=No                     \\
    ChgOffDate        & Date/Time & The date when a loan is declared to be in default    \\
    DisbursementDate  & Date/Time & Disbursement date                                    \\
    DisbursementGross & Currency  & Amount disbursed                                     \\
    BalanceGross      & Currency  & Gross amount outstanding                             \\
    MIS\_Status       & Text      & Loan status charged off=CHGOFF, Paid in full=PIF     \\
    ChgOffPrinGr      & Currency  & Charged-off amount                                   \\
    GrAppv            & Currency  & Gross amount of loan approved by bank                \\
    SBA\_Appv         & Currency  & SBA’s guaranteed amount of approved loan             \\ \midrule
    \end{tabular}
    \label{table1}
  \end{table}
\subsection{Analysis of variables}
\begin{figure}[htbp]
    \centering
    \includegraphics[width = 8cm]{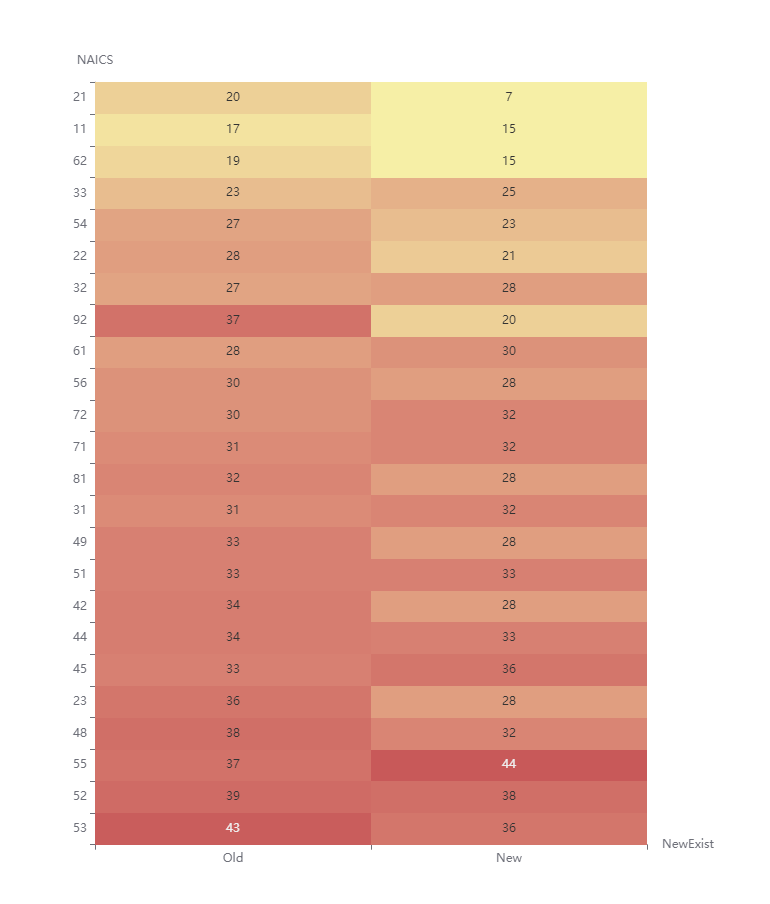}
    \caption{Default distribution based on feature \textit{NAICS} and \textit{NewExist}}
    \label{NAICS}
\end{figure}

We would like to identify some key variables before data training step. Intuitively, the gross amount of disbursement might be an important indicator of loan default. One possible explanation is that the larger the loan size, the more likely the underlying business will be established and expanded, thus increasing the likelihood of fulfilling the loan. Furthermore, for similar reasoning, the amount approved by the bank should be selected as well. The \textit{NAICS} (North American industry classification system code) number, the code for industrial classification, is expected to give out information. The first two digits in the code is an indicator of the industry the borrower belongs to. For example, “23” represents \textit{construction} and “52” \textit{finance and insurance}. It is expected that different industries can manifest a clear spectrum of average default rates and thus be a key indicator. Figure \ref{NAICS} shows such a distribution, where deeper color with large number in it means larger default rate.

\begin{figure}[htbp]
\centering
\begin{minipage}[t]{0.44\textwidth}
\centering
\includegraphics[width=7cm]{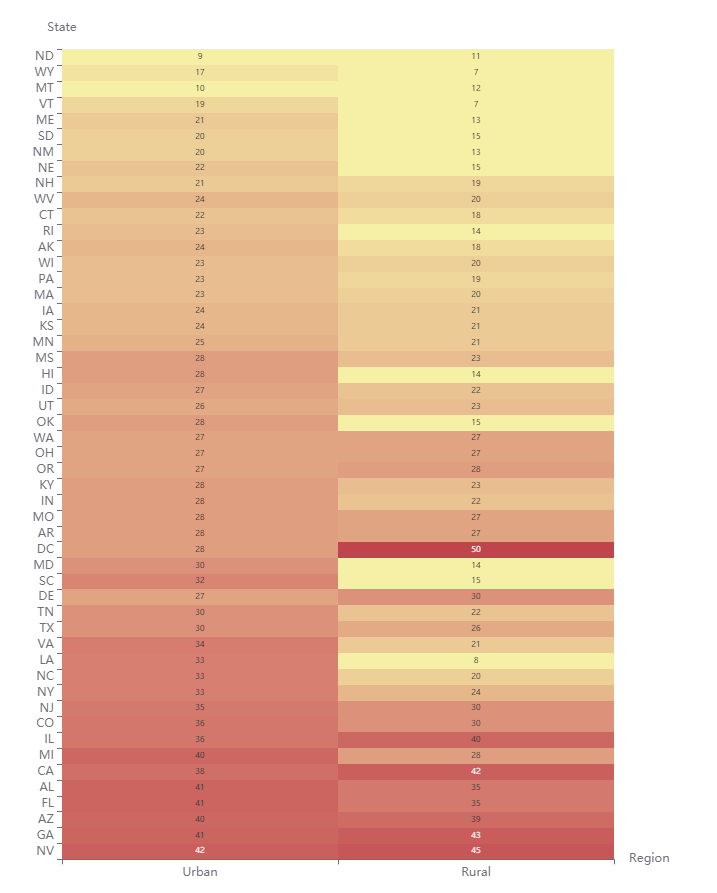}
\label{fig_12}
\caption{Default distribution based on feature \textit{State} and \textit{Region}}
\end{minipage}
\begin{minipage}[t]{0.44\textwidth}
\centering
\includegraphics[width=7cm]{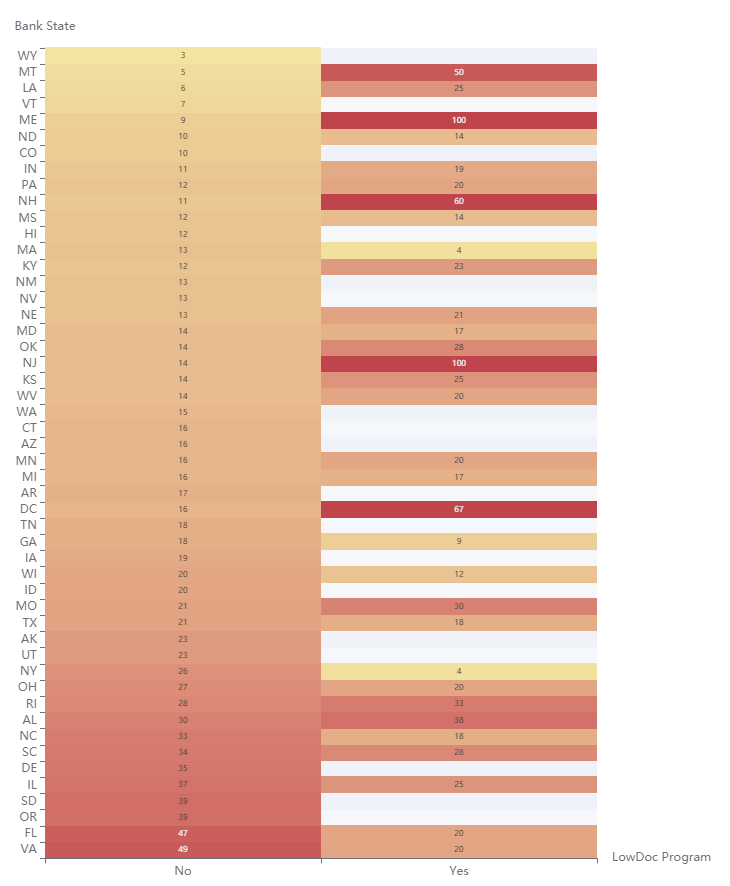}
\caption{Default distribution based on feature \textit{Bank State} and \textit{LowDoc}}
\label{fig_13}
\end{minipage}
\end{figure}

Other potential indicators are, for example, numbers of created jobs, whether to have a franchise, being urban or rural, whether to be newly established. One possible reasoning is that a company is more likely to default its loan if it is newly established, i.e. established for less than or equal to two years, because its scale is relatively small and its resistance to economic fluctuations is weak.

After visualization of several features, we have the following discoveries. First of all, contrary to our previous reasoning, the default rate of old businesses is slightly higher than that of new businesses (around  31.46\% versus 29.10\%). The default rate of urban borrowers is 32.08\%, significantly higher than that of rural companies (25.03\%). Even more surprising, companies attending \textit{LowDoc} Loan Program default at an average rate of 18.75\%, almost half of those who did not.

It is worth noticing that we can only intuitively perceive the connection between these features and default rates here. After setting up our model, we will illustrate the feature importance according to the model results and the connections will be quantified then.

\subsection{Processing of Variables}
There are categorical features that might serve as important variables. For CatBoost, these features can be directly put in the model and processed automatically. However, features like the State name, city of the borrowing company, bank name and NAICS code can not be directly processed by algorithms like logistic regression or SVM and should be converted to numerical features. The values these features take are not comparable to each other. Therefore, it is proper to convert the values into series of integers as we do in dummy variables. We decide to approach these features by their proportional relation with default rate. For example, if the borrower is from California State, then we calculate the proportion of borrowers from California that default the loan and create a new feature that indicates the overall underlying risk of companies from a particular State. We perform these techniques for features including borrower state, bank name, bank state and NAICS code. These constructed numerical features can then be inputted into the model along with other non-categorical features.

We can clearly see that performing conversions above can lose some underlying information of categorical variables. There is possibility that a categorical feature correlates with other features complicatedly and the relationship can only be properly represented if the model allows them to interact non-linearly. Converting to proportions saves only the information with default rates and misses the opportunity to consider the interplay above. CatBoost alleviates the issue by using combinations of categorical features as additional categorical features which capture high-order dependencies \citep{2017CatBoost}. However, the operations between features are too simple and combinations of non-categorical features cannot be captured. By performing synthetic features, we are expecting that more flexible combinations will be incorporated during model construction without losing computational feasibility.

\section{Methodology}
We used different machine learning algorithms to investigate the loan risk of small businesses and applied various models to predict the loan default rates. Different traditional machine learning algorithms and novel popular boosting ones were considered. CatBoost, which performs well when it comes to categorical variables, was our focus. To the best of our knowledge, there is still limitation for scenarios, like CatBoost, when applied in the comprehensive dataset which lacks enough continuous variables. To relax the limitation, we proposed a novel scheme by combining CatBoost with synthetic features.
\subsection{Logistic regression (LR)}
In the guidebook of the original dataset, Li et al. mentioned that Logistic regression is a powerful tool especially when coping with classification problems\citep{Li2018}. Logistic regression was introduced in the early 1970s \citep{cabrera1994logistic}, which aims to handle the limitations (such as linearity, normality, \textit{etc.}) of discriminant analysis. 
It received an extensive study in bankruptcy prediction \citep{2016Zieba}; \citep{jones2017predicting}; \citep{SON2019112816}; \citep{sigrist2019grabit}.  The relationship between probability of interest $P$ and a linear combination of predictors can be expressed as below:
$$
\log \left(\frac{P}{1-P}\right)=\sum_{i=1}^{K} \beta_{i} x_{i}+\beta_{0}
$$
After transformation, we can get $P$ as below:
$$
\begin{array}{l}
P=\frac{e^{\beta_{0}+\beta_{1} X_{1}+\beta_{2} X_{2}+\cdots+\beta_{K} X_{K}}}{1+e^{\beta_{0}+\beta_{1} X_{1}+\beta_{2} X_{2}+\cdots+\beta_{K} X_{K}}} \\
=\frac{1}{1+e^{-\left(\beta_{0}+\beta_{1} X_{1}+\beta_{2} X_{2}+\cdots+\beta_{K} X_{K}\right)}},
\end{array}
$$

\subsection{Support vector machine (SVM)}
The interruption of outliers will cause a substantial negative influence on the accuracy of the training model. In our exploration, there remain many outliers in the related dataset. So a kernel-based algorithm called support vector machine (SVM) had been thought to address this problem \citep{vapnik1995support}. We get the coefficients for SVM model by obtaining the solution of the minimization problem:
$$
{\arg \min \limits_{\beta_{0}, \beta}} \quad C \sum_{i=1}^{N} L_{\epsilon}\left(y_{i}-f(x_{i})\right) +\sum_{j=1}^{p} \beta_{j}^{2}
$$
where $L_{\epsilon}$ is a loss function and $C$ is the penalty.
We can get the solution by applying the sign function and a set of unknown parameters
$$
Z=f(y)=\textit{sign}\left(\sum_{i=1}^{N} y_{i} p_{i} K\left(x, x_{i}\right)+c\right)
$$
$\mathrm{p}_{\mathrm{i}}$ and $\mathrm{c}$ are used when constructing an optimal separating hyperplane, and $K\left(x, x_{i}\right)$ can be expressed as 
$$
K\left(\mathbf{x}, \mathbf{x}^{\prime}\right)=\exp \left(-\frac{\left\|\mathbf{x}-\mathbf{x}^{\prime}\right\|^{2}}{2 \sigma^{2}}\right)
$$
$\left\|\mathbf{x}-\mathbf{x}^{\prime}\right\|^{2}$ can be considered as the squared Euclidean distance, where $\mathbf{x}$ and $\mathbf{x}^{\prime}$ are the two feature vectors. $\sigma$ can be transferred to a new parameter $\gamma=\frac{1}{2 \sigma^{2}}$ using an equivalent definition 
$K\left(\mathbf{x}, \mathbf{x}^{\prime}\right)=\exp \left(-\gamma\left\|\mathbf{x}-\mathbf{x}^{\prime}\right\|^{2}\right)$

When we applied SVM model to our dataset, we found that it was a delicate task to tune the related parameter and it would be time-consuming to get the whole model. Furthermore, the SVM model did not meet our expectation of a good model, which just requires a low memory capacity.
\subsection{Random forest (RF)} 
Random forest is a popular technique to eliminate the barrier of high correlations between single tree in bagging process \citep{breiman2001random}. It is a rule-based algorithm commonly used in financial application to refine accuracy of the predicting. It uses a procedure called "feature bagging", where every tree is regarded as a supply of single vote, appointing each variable to the most possible class of output. We get the output function by using the probability distribution $\mathrm{p}_{\mathrm{t}}(\mathrm{y} \mid \mathrm{x})$ of each tree:
$$
Z = \arg \max{\frac{1}{T} \sum_{t=1}^{T} p_{t}(y / x)}
$$

In applying to the $\mathrm{RF}$ model in the related dataset, we found it performs well in coping with noisy and complex data. Furthermore, it had a good accuracy of forecasting, compared with many traditional modelling scenarios as mentioned above.

\subsection{MLPclassifier}
\begin{figure}[htbp]
  \centering 
  \includegraphics[width = 13.3cm]{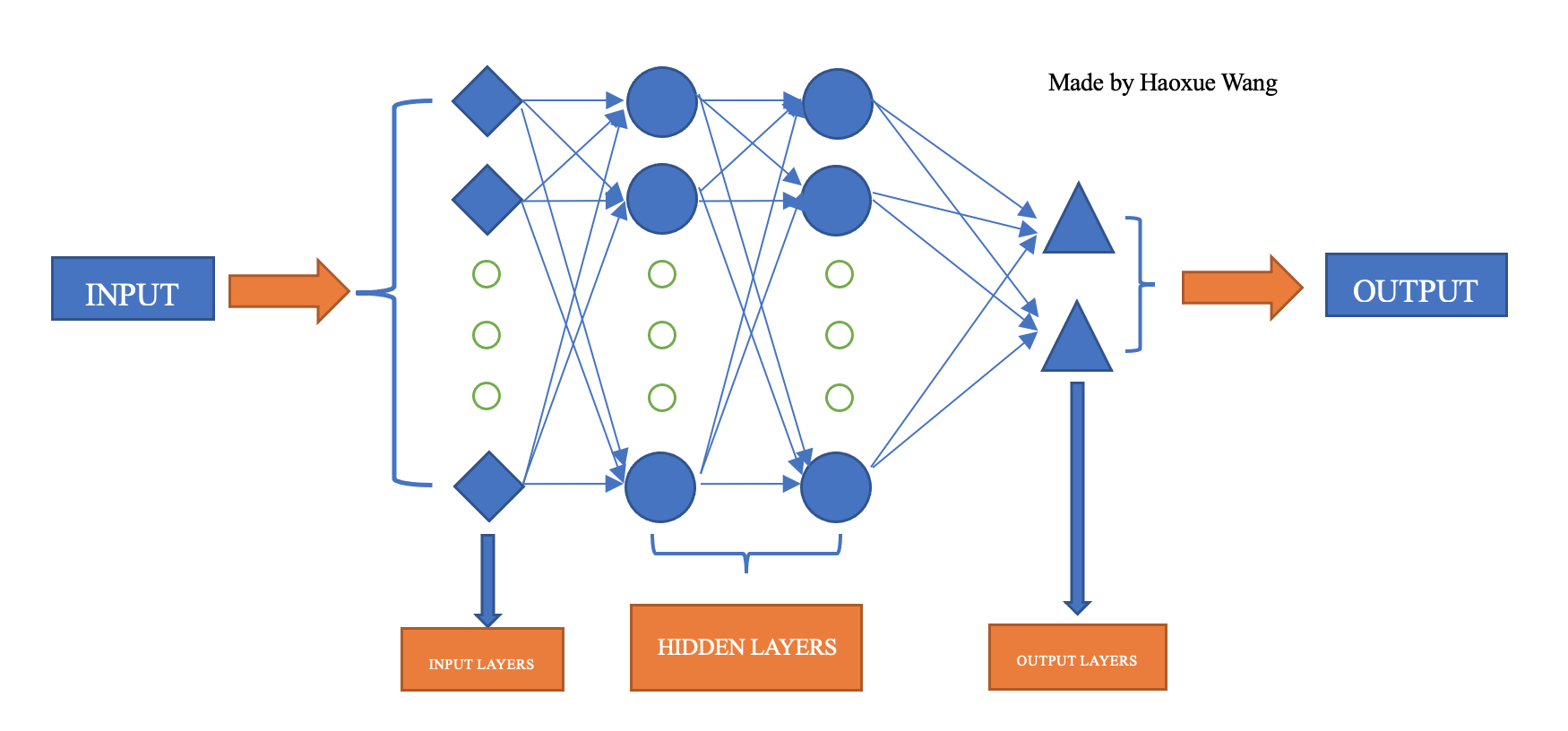}
  \caption{The structure of MLPclassifier}
  \label{mlp}
\end{figure}

Multilayer perceptron (MLP) has been a popular artificial neural network model since it was proposed, which establishes non-linear mappings between sets of input data onto a set of sound outputs. The difference of each layer lies in the setting number of nodes and various activation functions. In most cases, Sigmoid function is used in the hidden units as an activation function. Typically, for better training of deeper networks, Relu activation function can be an alternative choice. After each iteration, activation functions are passed only from one layer. In applications, it enjoys statistical benefits and unique properties in diverse fields. For example, Ahad et al. had employed MLP in speech recognition and showed its talent to recognize digits from zero to nine \citep{Ahad}. Besides, MLP has gained popularity in image recognition and machine translation software as well. The structure of MLPclassifier is shown in figure \ref{mlp}

\subsection{Gradient Boosting Machine and XGBoost}
GBDT (Gradient Boosting Decision Tree) is a powerful model in machine learning, which combines weak classifiers (decision trees) for iterative training to attain the optimization of the previous models. This kind of algorithm is of good training performance and enjoys excellent benefits as a solution to overfit. GBDT has been widely used in the industry, which shows outstanding properties when solving tasks such as multi-category, click-through rate prediction, search sorting, \textit{etc.}; It also has been a practical tool in competitions that called for the high accuracy of data predicting. The following presents a structure of the generic gradient tree-boosting algorithm. Different algorithms are obtained by applying different loss functions, by optimizing which we solve the overfitting problem. To the best of our knowledge, when we have a custom base-learner $h(x, \theta)$ (such as decision tree), and a loss function $\psi(y, f(x)) ;$ Estimating the parameters directly can be an impossible task, as a consequence, an iterative process was proposed to solve this problem. The training model will be continued to keep updated by selecting a new base-learner function $h\left(x, \theta_{t}\right)$, we can get the function as below:
$$
g_{t}(x)=E_{y}\left[\frac{\partial \psi(y, f(x))}{\partial f(x)} \mid x\right]_{f(x)=\tilde{f}^{t-1}(x)}
$$
By transforming this function, we can replace the original complex optimization task with least-square optimization problem we are familiar with:
$$
\left(\rho_{t}, \theta_{t}\right)=\arg \min _{\rho, \theta} \sum_{i=1}^{N}\left[-g_{t}\left(x_{i}\right)+\rho h\left(x_{i}, \theta\right)\right]^{2}
$$

\begin{algorithm}
  \caption{Gradient Boost}\label{algorithm1}
  \KwData{Firstly we have input $(x, y)_{i=1}^{N}$; 
   choose number of iterations $M$; 
   decide the loss-function $\Psi(y, f)$; 
   set the base-learner model $h(x, \theta)$
   set up $\widehat{f}_{0}$ }
   
   \For{$t=1$ to $M$} {
 calculate $g_{t}(x)$
 
 fit a regression tree to target another new function $h\left(x, \theta_{t}\right)$
 
 compute the best gradient descent step-size $\rho_{t}$ :
 $$
 \rho_{t}=\arg \min _{\rho} \sum_{i=1}^{N} \Psi\left[y_{i}, \widehat{f}_{t-1}\left(x_{i}\right)+\rho h\left(x_{i}, \theta_{t}\right)\right]
 $$
 
 update:
 $\widehat{f}_{t} \leftarrow \widehat{f}_{t-1}+\rho_{t} h\left(x, \theta_{t}\right)$
 }
 
 \end{algorithm}

 eXtreme Gradient Boosting (XGBoost) owns innovative improvement compared with previous gradient boosting
 algorithm. It is a highly practical, pliant and variable tool that was developed by Chen and Guestrin\citep{chen2016xgboost} and pushed the gradient boosting to the spotlight. We can tell the main advantage among 
 XGBoost by using another new more regularized technique for handling the overfitting problem and achieve better accuracy. In the process of applying the model to our dataset, we found it reacted faster and became more robust when tuning the model parameter. We can regard the regularization process as adding a new term to the
 loss function, as: 
 $$
 L(f)=\sum_{i=1}^{n} L\left(\hat{y}_{i}, y_{i}\right)+\sum_{m=1}^{M} \Omega\left(\delta_{m}\right)
 $$
 with
 $$
 \Omega(\delta)=\alpha|\delta|+0.5 \beta\|w\|^{2}
 $$
 where $|\delta|$ is the number of branches, $w$ is the value of each leaf and $\Omega$ is the regularization function. XGBoost uses a new gain function, as:
 $$
 \begin{array}{c}
 G_{j}=\sum_{i \in I_{j}} g_{i} \\
 H_{j}=\sum_{i \in I_{j}} h_{i} \\
 \textit { Gain }=\frac{1}{2}\left[\frac{G_{L}^{2}}{H_{L}+\beta}+\frac{G_{R}^{2}}{H_{R}+\beta}-\frac{\left(G_{R}+G_{L}\right)^{2}}{H_{R}+H_{L}+\beta}\right]-\alpha
 \end{array}
 $$
 where
 $$
 g_{i}=\partial_{\hat{y}_{i}} L\left(\hat{y}_{i}, y_{i}\right)
 $$
 and
 $$
 h_{i}=\partial_{\hat{y}_{i}}^{2} L\left(\hat{y}_{i}, y_{i}\right)
 $$
  The result of XGBoost is an altogether different function from the beginning, as the ending function
 is the combination of various functions.

\subsection{LightGBM}
LightGBM (Light Gradient Boosting Machine) can be regarded as an improvement of the previous model based on the framework of GBDT algorithm. One main difference of LightGBM is that it supports high-efficiency parallel training but does not sacrifice training speed. As its name indicated, it enjoys a property which calls for lower memory consumption, as a consequence, it has been used in many circumstances especially for quick processing of massive data\citep{ke2017lightgbm}. It is a histogram-based algorithm developed by Microsoft. Furthermore, a leaf-wise strategy is used in LightGBM to develop different trees and do the split part by gaining the largest variance. 

\begin{figure}[htbp]
  \centering 
  \includegraphics[width = 8.3cm]{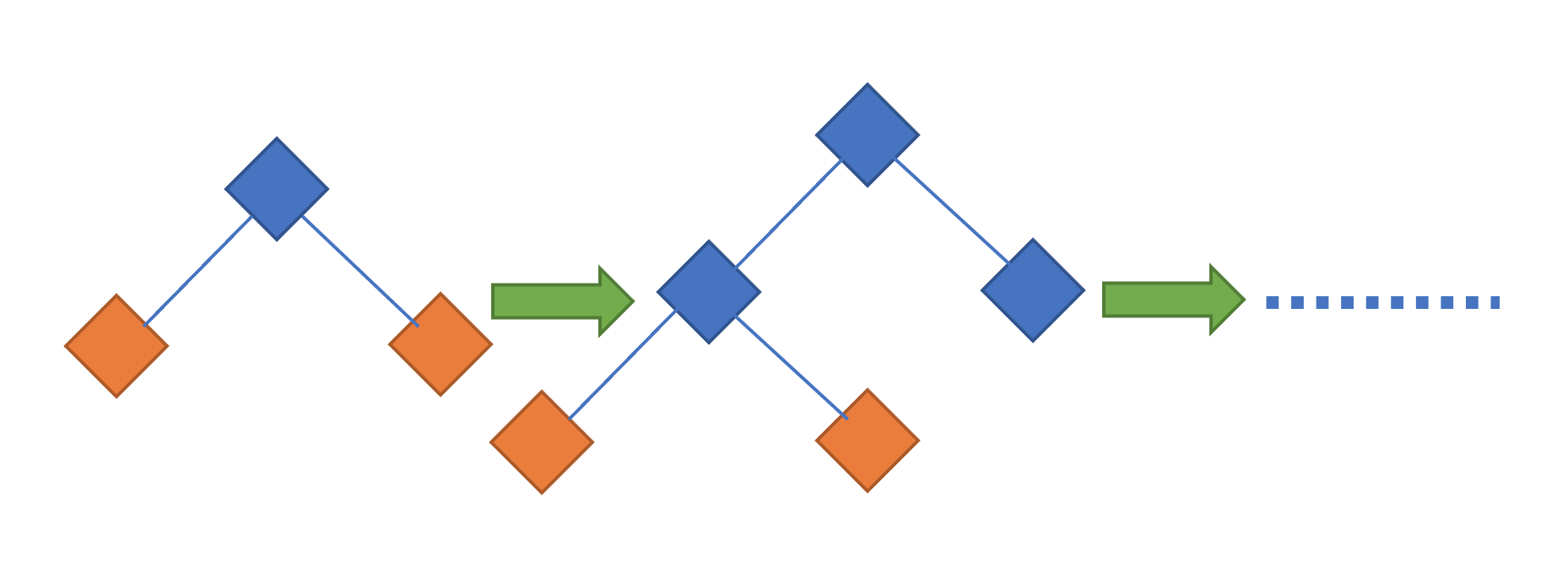}
  \caption{XGBoost Level-wise tree growth}
  \label{xgboost}
\end{figure}

\begin{figure}[htbp]
  \centering 
  \includegraphics[width = 10.3cm]{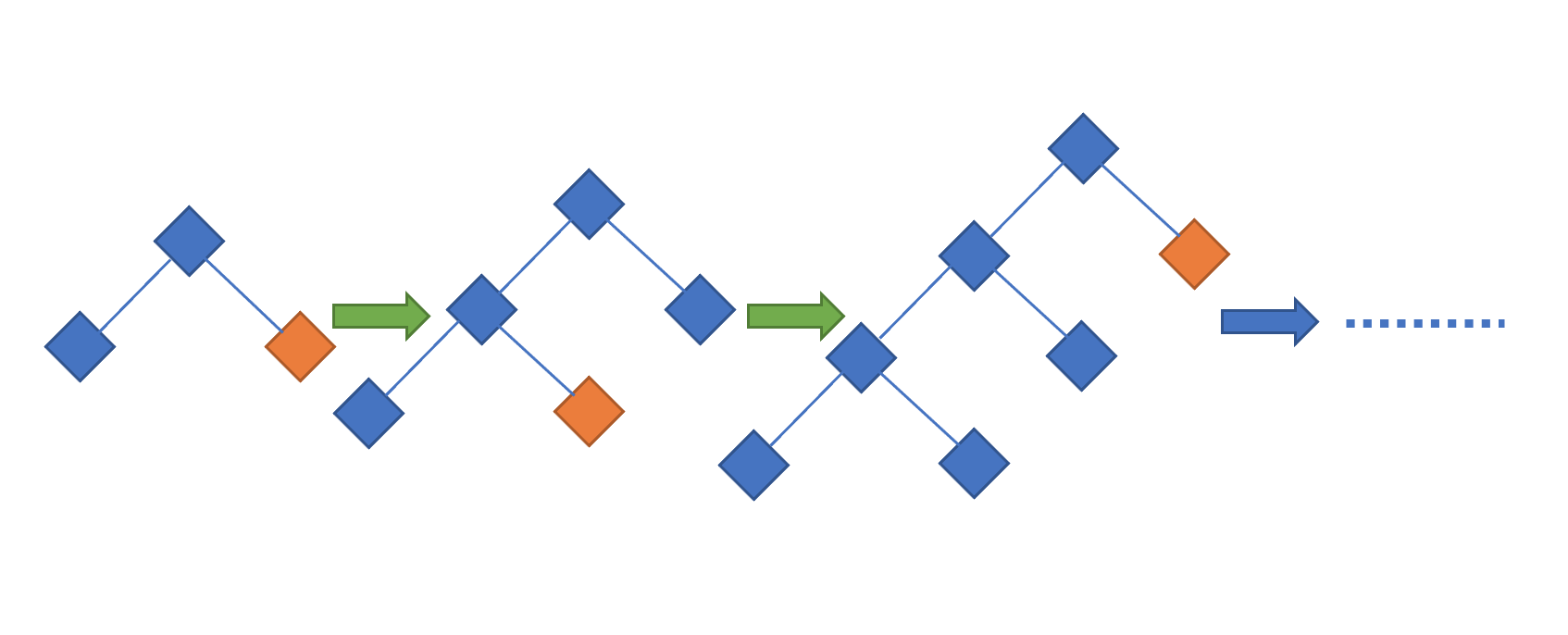}
  \caption{LightGBM Leaf-wise tree growth}
  \label{light}
\end{figure}

A variety of literature to cope with different problems have been done via LightGBM, they all based on the following function to gain estimated variance $V_{j}^{*}(d)$ over the subset $A \cup B$.

\begin{align*}
V_{j}^{*}(d)=& \frac{1}{n}\left(\frac{\left(\sum_{x_{i} \in A_{l}} g_{i}+\frac{1-a}{b} \sum_{x_{i} \in B_{l}} g_{i}\right)^{2}}{n_{l}^{j}(d)}\right.\\
&\left.+\frac{\left(\sum_{x_{i} \in A_{r}} g_{i}+\frac{1-a}{b} \sum_{x_{i} \in B_{r}} g_{i}\right)^{2}}{n_{r}^{j}(d)}\right)
\end{align*}

where $A_{l}=\left\{x_{i} \in A: x_{i j} \le d\right\}, A_{r}=\left\{x_{i} \in A: x_{i j}>d\right\}$,
$B_{l}=\left\{x_{i} \in B: x_{i j} \le d\right\}, B_{r}=\left\{x_{i} \in B: x_{i j}>d\right\}, d$ is
the point in dataset in which the split is computed when finding the best variance performance.
The main difference between LightGBM and XGBoost is shown as the figures \ref{xgboost} and \ref{light}

\subsection{CatBoost}
CatBoost, which is the abbreviation for "categorical boosting", concentrates on categorical features. It is a gradient boosting decision tree (GBDT) algorithm as well. It is introduced by Prokhorenkova et al. and Dorogush et al. to obtain the lowest loss when it comes to many categorical features in the dataset. \citep{dorogush2018catboost};\citep{prokhorenkova2017catboost}However, we can find some differences between it and traditional GBDT algorithms.
\begin{itemize}
  \item CatBoost can be more effective as it deals with categorical features during training time, compared to many other algorithms done in preprocessing time. It transformed categorical features into numeric ones by adding some hyperparameters.
  \item To the best of our knowledge, CatBoost performs well when applied in small size dataset as the whole dataset can be used.\citep{prokhorenkova2017catboost}
  \item Target statistics(TS) is the process to improve the efficiency, which substitutes the original category $x^i_k$ of k-th training variables with one number feature equivalent to some target statistic (TS)  $\hat{x}^i_k$. we can obtain $\hat{x}_{k}^{i} \approx \mathbb{E}\left(y \mid x^{i}=x_{k}^{i}\right)$. 
  \item Moreover, a random permutation of the dataset is implemented when choosing the tree structure, which calculates an average of leaf values. If a permutation is
  $\Theta=\left[\sigma_{1}, \cdots, \sigma_{n}\right]_{n}^{T},$ it is replaced with
  
\begin{center}
  $x_{\sigma_{p, k}}=\frac{\sum_{j=1}^{p-1}\left[x_{\sigma_{j, k}}=x_{\sigma_{p, k}}\right] \cdot Y_{\sigma_{j}}+\beta \cdot P}{\sum_{j=1}^{p-1}\left[x_{\sigma_{j, k}}=x_{\sigma_{p, k}}\right]+\beta}$
\end{center}
  After transforming the equation, we can get the estimated $\mathbb{E}\left(y \mid x^{i}=x_{k}^{i}\right)$ by making use of the average value of $y$ over the dataset for training with the same category $x^i_k$. 

  \item A greedy way is thought in the assemblage of all the categorical features to new ones when developing a new split for the tree. There is no combination process occurring when doing the first split of the tree. Instead, it is applied for the second and the subsequent splits.
  \item When applying an unbiased boosting dealing with categorical features, TS method is used to transfer categorical features into number values, at the same time, the distribution will change, which shows a different property from the original distribution. As a consequence, the resulted deviation of the solution will be a problem to be solved, which is inevitable when we talk about traditional GBDT methods. Nevertheless, for CatBoost, Prokhorenkova et al. introduced a novel approach to handle the limitation and provided a theoretical analysis. The novel approach, which was called ordered boosting, can be expressed to the pseudo-code as shown as algorithm \ref{algorithm2}:

\begin{algorithm}
\caption{Ordered boosting}\label{algorithm2}
\KwData{$\{\left.\left(X_{k}, Y_{k}\right)\right\}_{k=1}^{n}$ ordered according to $\sigma,$ the number of trees $I$;}
$\sigma \leftarrow$ random permutation of $[1, n]$; 

$M_{i} \leftarrow 0$ for $i=1, \cdots, n$;

  \For{$t \leftarrow 1$ to $I$}{
  \For{$i \leftarrow 1$ to $n$}{  
    $r_{i} \leftarrow y_{i}-M_{\sigma(i)-1}\left(X_{i}\right)$;
  }  
  \For{$i \leftarrow 1$ to $n$}{ 
  $\Delta M \leftarrow$ LearnModel $\left[\left(X_{i}, r_{j}\right): \sigma(j) \leq i\right]$;
  
  $M_{i} \leftarrow M_{i}+\Delta M$;
  }
}
\end{algorithm}

\item In CatBoost, we often apply stochastic permutations of the training data, which make a difference to increase the capacity of the algorithm’s robustness. As different permutations will be considered in the model, CatBoost will not contribute to an overfitting problem. When we have each permutation $\sigma, \mathrm{n}$ different models $M_{i}$ will be trained as shown above, indicating that when building a tree, there is a requirement to keep calculating $O\left(n^{2}\right)$ approximations for each permutation $\sigma$ again and update each model $M_{i}$ from $M_{i}\left(X_{1}\right)$ to $M_{i}\left(X_{i}\right)$. Moreover, the complexity of this process can be expressed as $O\left(s \times n^{2}\right)$. When we apply it to our dataset, we consider a beneficial trick which decreases the complexity of one tree construction to $O(s \times n)$. Precisely speaking, when we handle individual permutation, rather than maintaining and developing $O\left(n^{2}\right)$ values $M_{i}\left(X_{i}\right),$ we choose to keep values $M_{i}^{\prime}\left(X_{j}\right)$, $i=1, \cdots,\left[\log _{2}(n)\right], j<2^{i+1}$, where $M_{i}^{\prime}\left(X_{j}\right)$ can be regarded as the approximation for the same $j$ on the basis of the first $2^{i}$ samples. As a consequence, the numeric predictions value $M_{i}^{\prime}\left(X_{j}\right)$ will kept smaller than $\sum_{0 \le i \le \log _{2}(n)} 2^{i+1}<4 n$.

The structure of the CatBoost algorithm is shown in fugure \ref{fig:figure1label4}.
\begin{figure}[htbp]
\centering 
\includegraphics[width = 13.3cm]{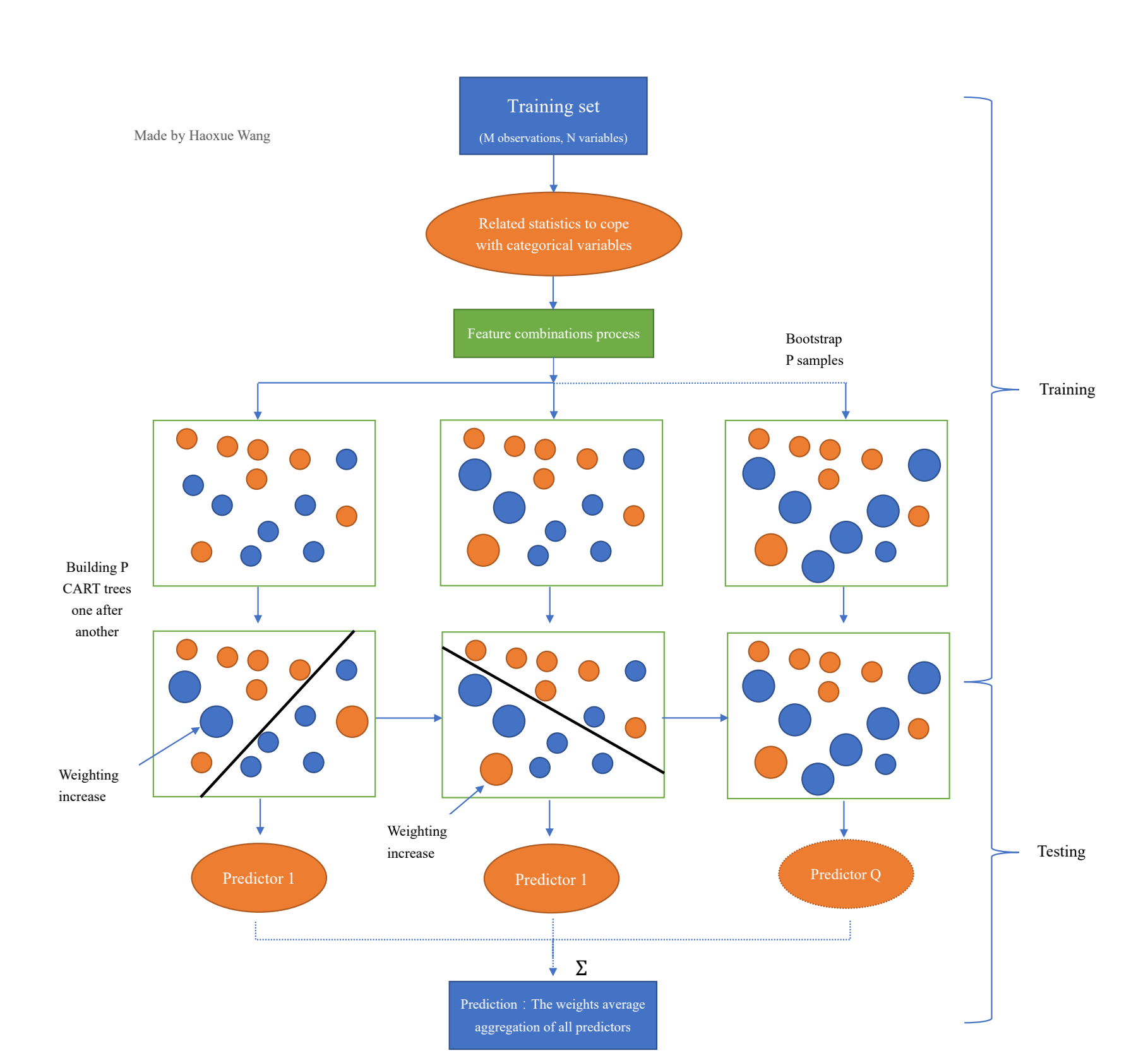}
\caption{The structure of CatBoost}
\label{fig:figure1label4}
\end{figure}
\end{itemize}

\section{CatBoost with synthetic feature}
When we explored the dataset, we found that the size of features in the dataset is small-scale. Consequently, we came up with a novel technique called synthetic feature to generate features based on the continuous feature importance. It is an effective way to make use of the data as CatBoost is a tree-based model. When applied to the data, there are two ways to realize this approach. It will be easier if we just select randomly two continuous variables and make simple calculations to get new features. However, in order to make the generation process more effective and less time-consuming. We first make a rank of features according to their contribution to the classification result. Using the combined first two ranked features, we perform a random arithmetical operation on them. In order to predict the probability of selecting the seed features, we consider the popularity of that feature in the created forest. Then we describe the popularity of the feature in the forest using the total numeric value of occurrences in trees made up of the forest. 

After the process mentioned above, we can make use of the most popular features to continue the reproduction process. This kind of technique to generate new features can be regarded as an innovative way to filtrate the less important features and amplification the influence of strongest feature, which plays an important role in the accuracy of the model.

We can combine different arithmetic operations for the selected features, $\{+,-, *, /\}$ are the commonly used when we do the calculation. For extension, we can undergo different arithmetic operations at the same time.

\begin{algorithm}
  \caption{CatBoost with synthetic features}\label{algorithm}
  \KwData{Input $: \mathcal{D}:$ training input, $D_{\text {new }}:$ sets with synthetic features, $L:$ the amount of base learners, $\eta:$ threshold for features incorporating, $F:$ feature importance;\\
  \qquad Output:$H=\left\{h_{1},\ldots, h_{K}\right\}:$ set of ending learners}
  \For{$k=1, \ldots, K$}{
    Train $h_{k}$ learner based on $\mathcal{D}$; 
    
    Remove features from $\mathcal{D}$ for which performance $p_{d}<\eta$;
    
    Accept model $h_{k}$ if $p_{d}>\eta$;
    
    \For{$d=1, \ldots, D_{\text {new }}$}{ 
    Sample features $f_{1}$ and $f_{2}$ are obtained based on the feature importance $F$; 
    
    Sample operation is processed from $\{+,-, *, /\}$ Generate new feature $f_{\text {new }}=f_{1} \circ f_{2}$; 
    
    Increasing sets $\mathcal{D}$ with new features of $f_{\text {new }}$; 
    }
  }
  \Return{$H=\left\{h_{1}, \ldots, h_{K}\right\}$} 
  \end{algorithm}

\section{Model Setup}
In this section, we clarify specific model settings and important details during training of data. As measures of predictive power of the different machine learning models, we focus on two evaluation metrics: accuracy and the area under the ROC curve (AUC).

\subsection{Tuning parameters}
For each machine learning model considered in this paper, we examine the performance of different settings of training parameters using ten folds cross-validation methodology. In Appendix \ref{tuning results} we present only the best set of parameters and the corresponding performance for each type of the considered classifiers.

\subsection{Incorporate Synthetic features}
In order for our model to capture features in a more flexible way, we create synthetic features heuristically based on the feature importance given by the construction of trees.

The whole procedure of CatBoost with synthetic features is described in Algorithm \ref{algorithm}. In each iteration that trains a base learner using dataset $D$, features with importance, which is measured by its frequency of appearance, lower than a given threshold are removed. Then the model constructs a new sampling distribution $\theta_F$ according to base learner $h_k$. Further evaluation of the features' popularity is carried out using the learned distribution.

Also, in each iteration, the synthetic features are generated based on the following framework. Two features $f_1$ and $f_2$, either the original features or synthetic ones generated previously, are sampled from distribution $\theta_F$. Then the operation $\circ$ is sampled from the set $\{+, -, *, /\}$ using uniform distribution and the value of new feature $f_{new} = f_1 \circ f_2$ is assigned accordingly. The process of generating synthetic features above is repeated until the desired number of synthetic features, $D_{new}$, is reached. By incorporating new features, the extended dataset is further used to construct the $h_{k+1}$ base model in the next iteration.

\section{Results and Discussions}

\subsection{Model Evaluation}

\begin{table}[ht]
\caption{Result of some popular machine learning algorithms}
\centering
\label{machine learning}
\begin{tabular}{ccccc}\midrule
Model    & Logistic Regression                  & SVM        & Random Forest    & MLPclassifier \\
Accuracy & 84.39\%             & 91.42\%    & 95.53\%          & 92.91\%       \\
AUC      & 90.09\%             & 96.21\%    & 98.52\%          & 97.26\%       
\end{tabular}
\end{table}

\begin{table}[ht]
\caption{Result of Boosting-type machine learning algorithms}
\centering
\label{ensemble}
\begin{tabular}{ccccc}\midrule
Model     & lightGBM      & XGBoost    & CatBoost    & CatBoost with synthetic features\\
Accuracy  &  94.74\%      & 95.56\%    & 95.74\%     & 95.84\%   \\
AUC       &  98.15\%      & 98.53\%    & 98.59\%     & 98.80\%
\end{tabular}
\end{table}

Based on the best results of parameter tuning, we run the dataset \textit{national SBA} on each model and evaluate the average performance. The results, including accuracy and AUC, are listed in table \ref{machine learning} and table \ref{ensemble}. 

\begin{figure}[htbp]
  \centering 
  \includegraphics[width = 8.3cm]{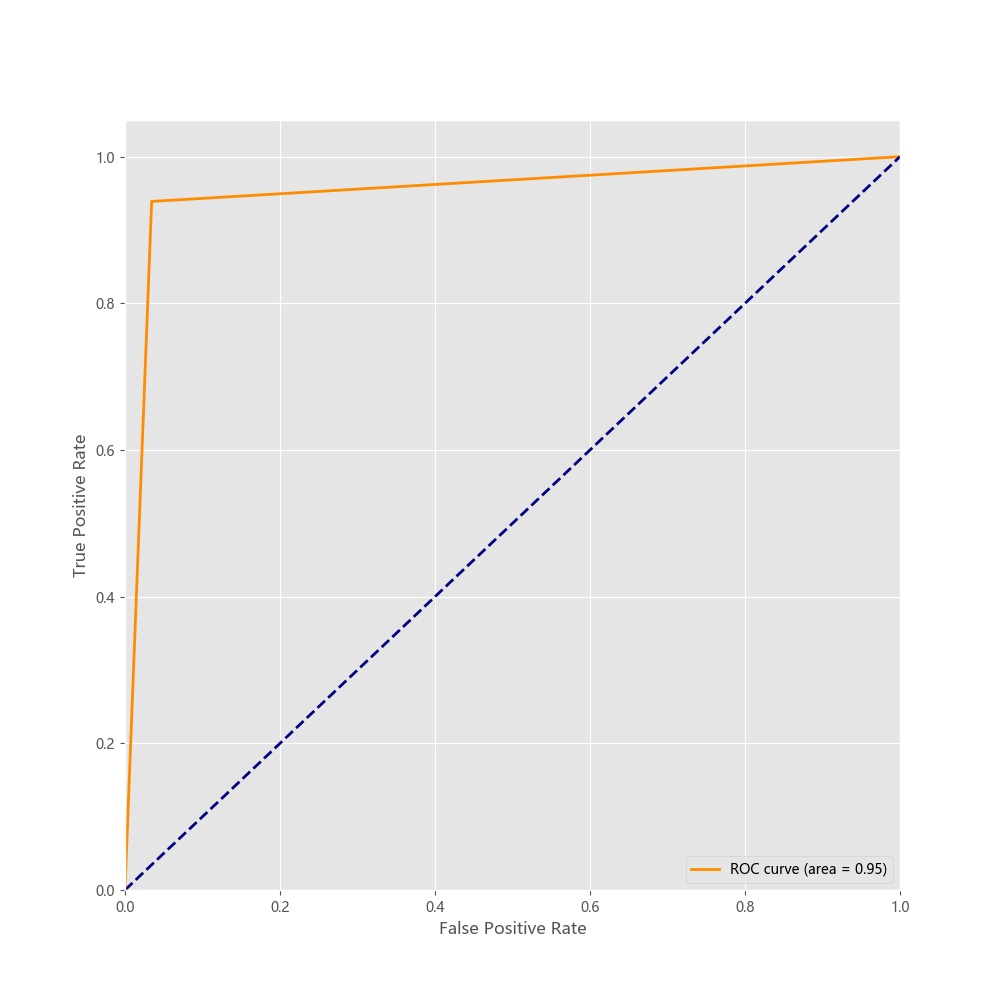}
  \caption{The ROC curve of CatBoost}
  \label{fig:figure1label5}
  \end{figure}
  
\begin{figure}[htbp]
  \centering
  \begin{minipage}[t]{0.48\textwidth}
  \centering
  \includegraphics[width=6cm]{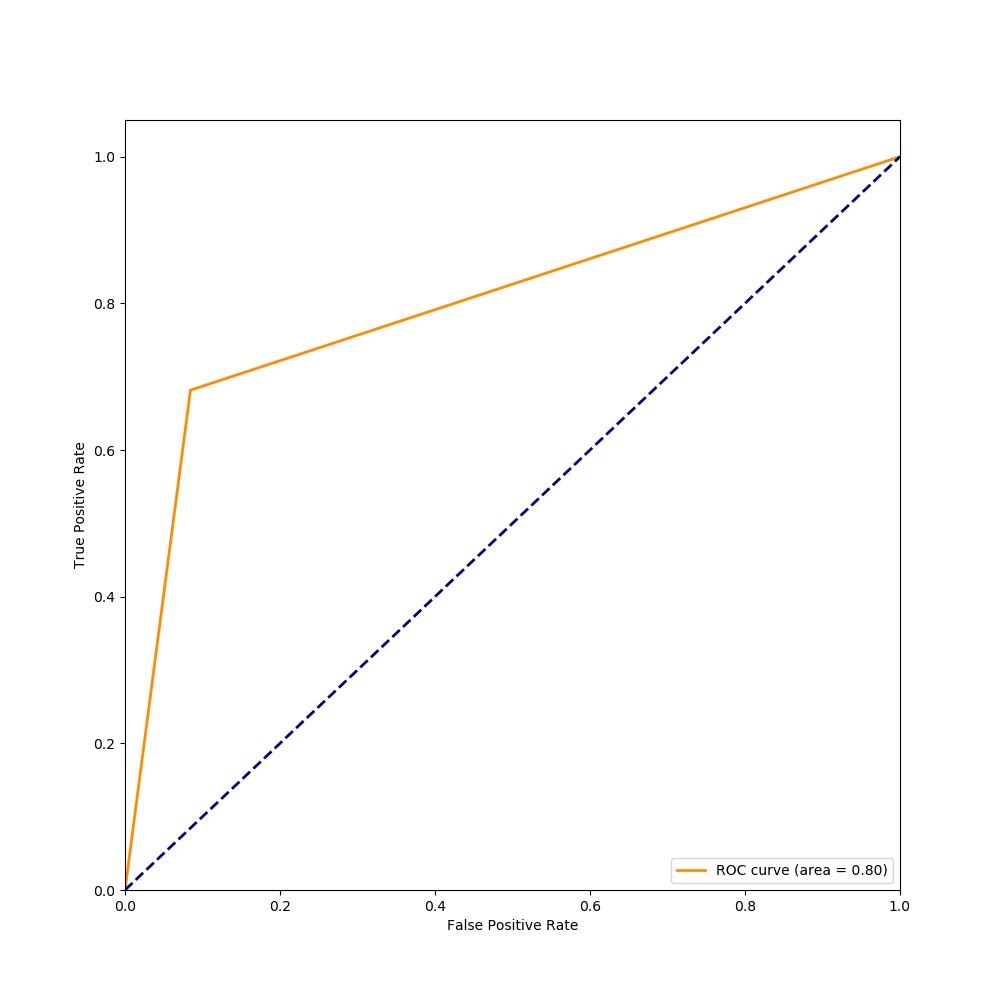}
  \caption{The ROC curve of Logistic Regression}
  \end{minipage}
  \begin{minipage}[t]{0.48\textwidth}
  \centering
  \includegraphics[width=6cm]{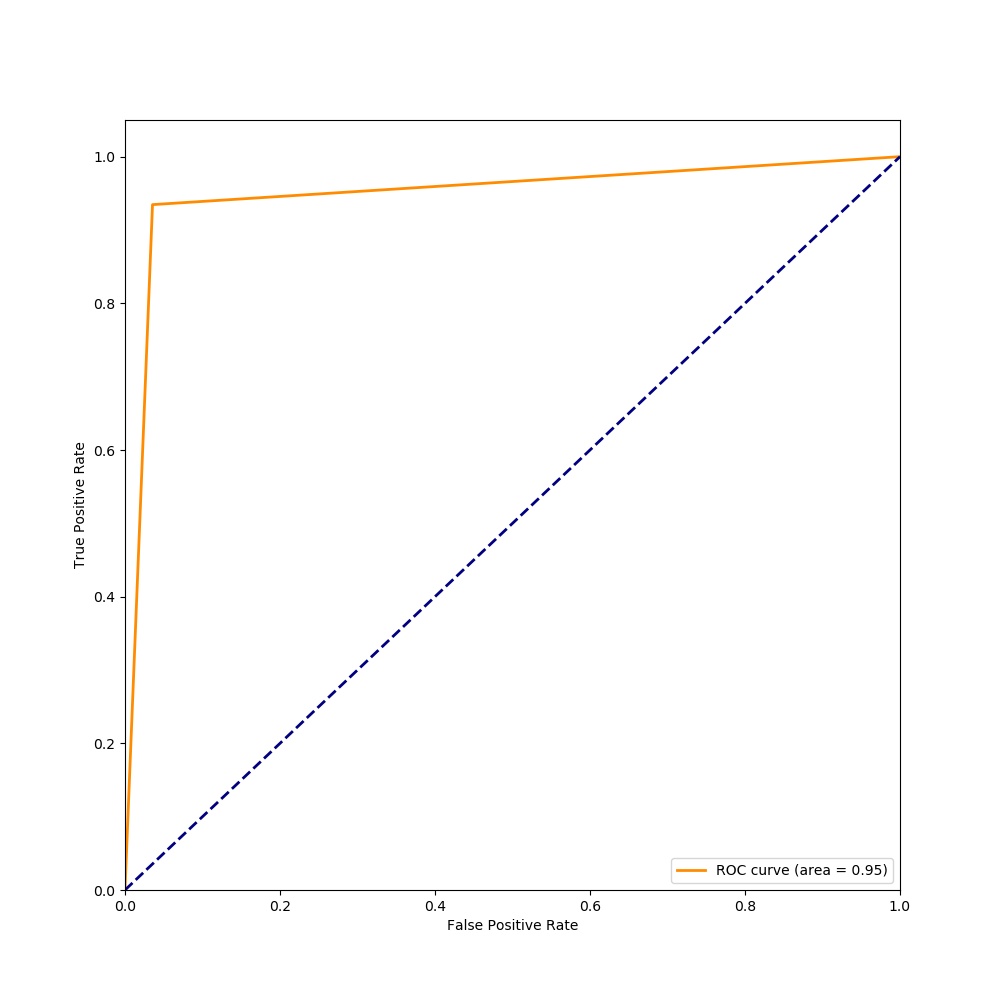}
  \caption{The ROC curve of Random Forest}
  \end{minipage}
  \end{figure}

\begin{figure}[htbp]
\centering
    \begin{minipage}[t]{0.48\textwidth}
    \centering
    \includegraphics[width=6cm]{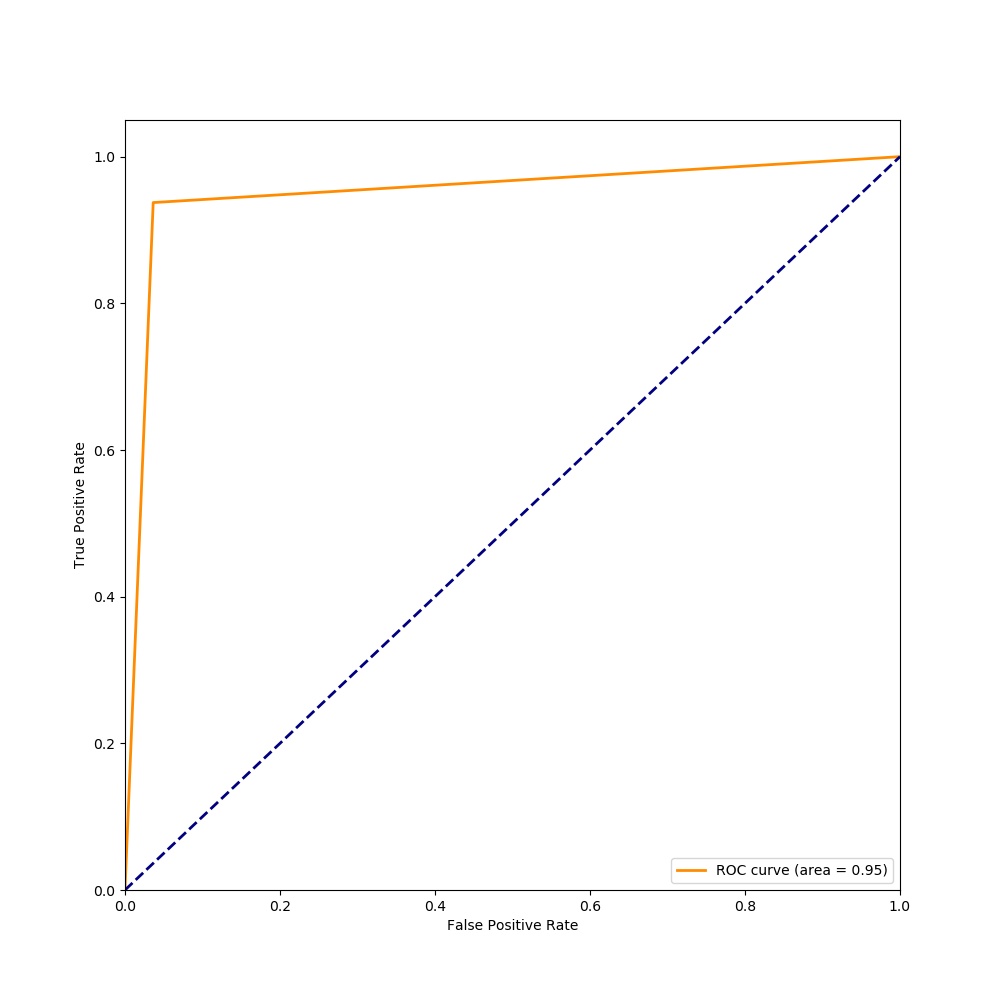}
    \caption{The ROC curve of XGBoost}
    \end{minipage}
    \begin{minipage}[t]{0.48\textwidth}
    \centering
    \includegraphics[width=6cm]{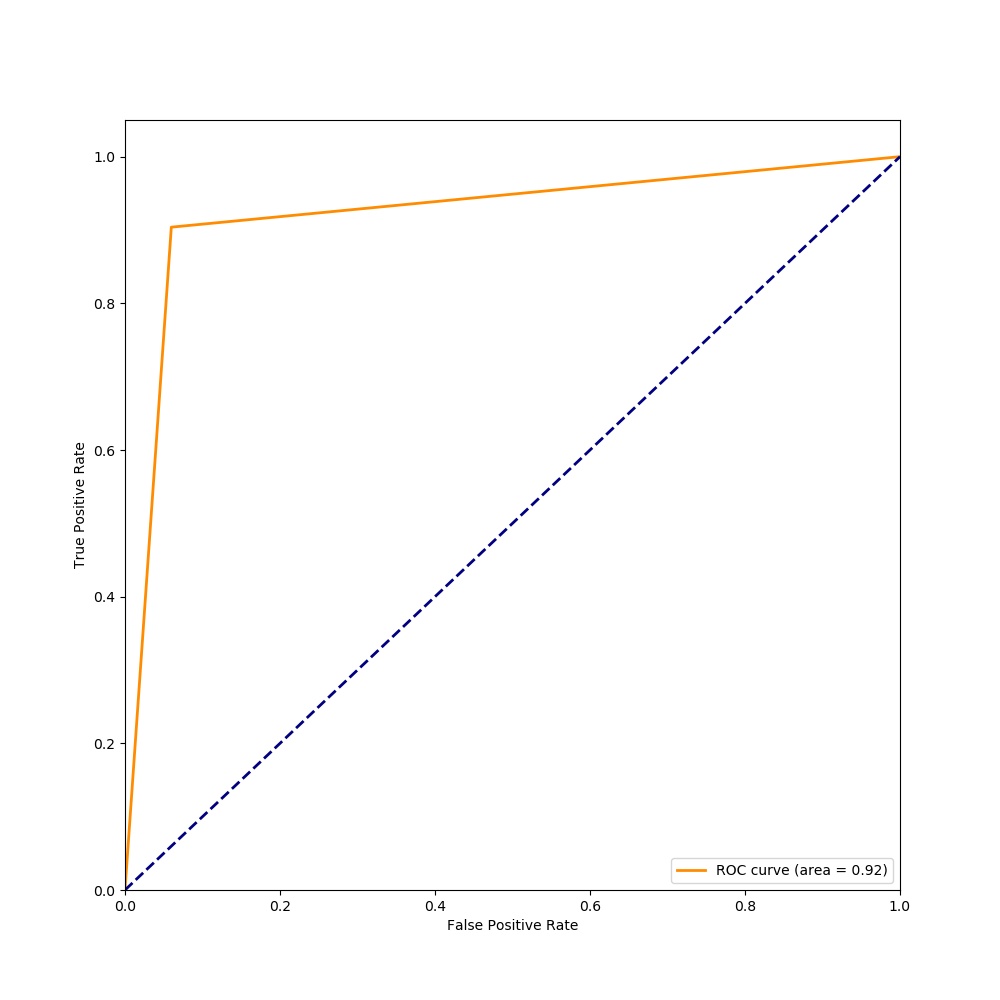}
    \caption{The ROC curve of KNN}
    \end{minipage}
    \end{figure}

    \begin{figure}[htbp]    
    \centering
    \begin{minipage}[t]{0.48\textwidth}
    \centering
    \includegraphics[width=6cm]{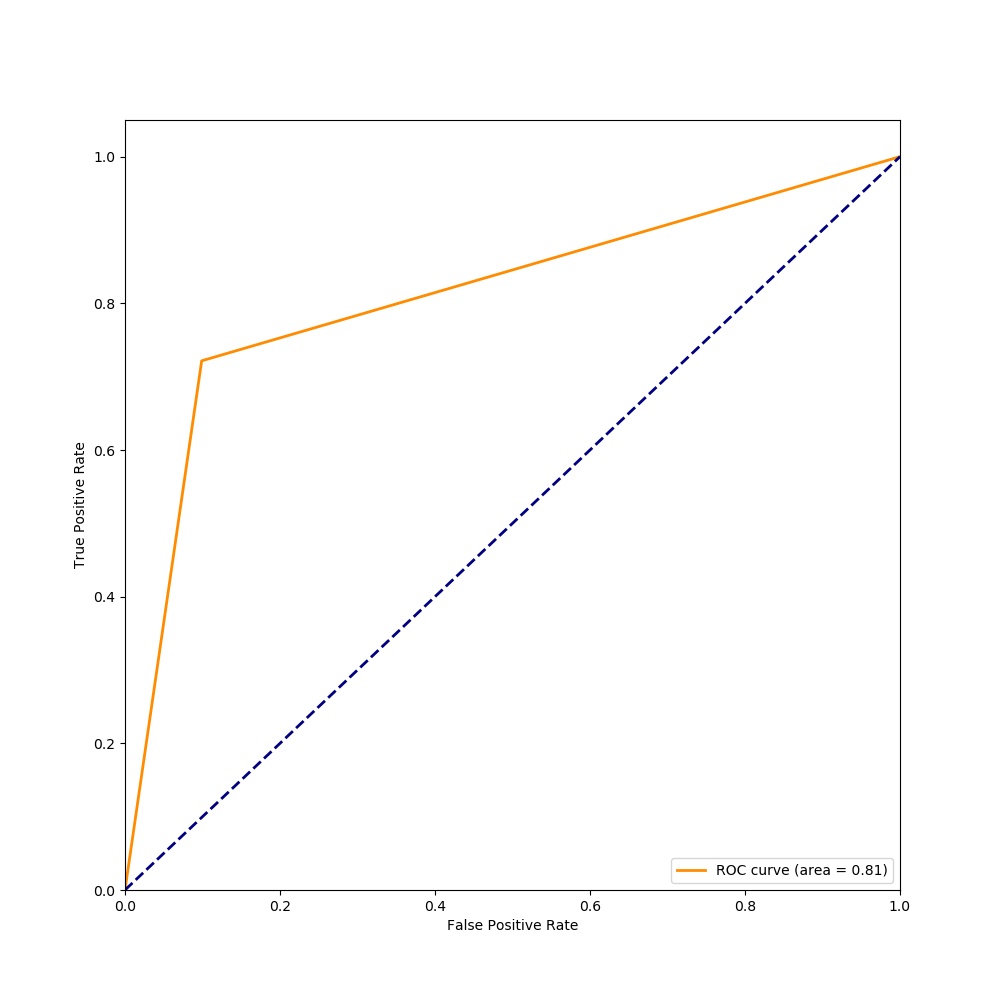}
    \caption{The ROC curve of SVM}
    \end{minipage}
    \begin{minipage}[t]{0.48\textwidth}
    \centering
    \includegraphics[width=6cm]{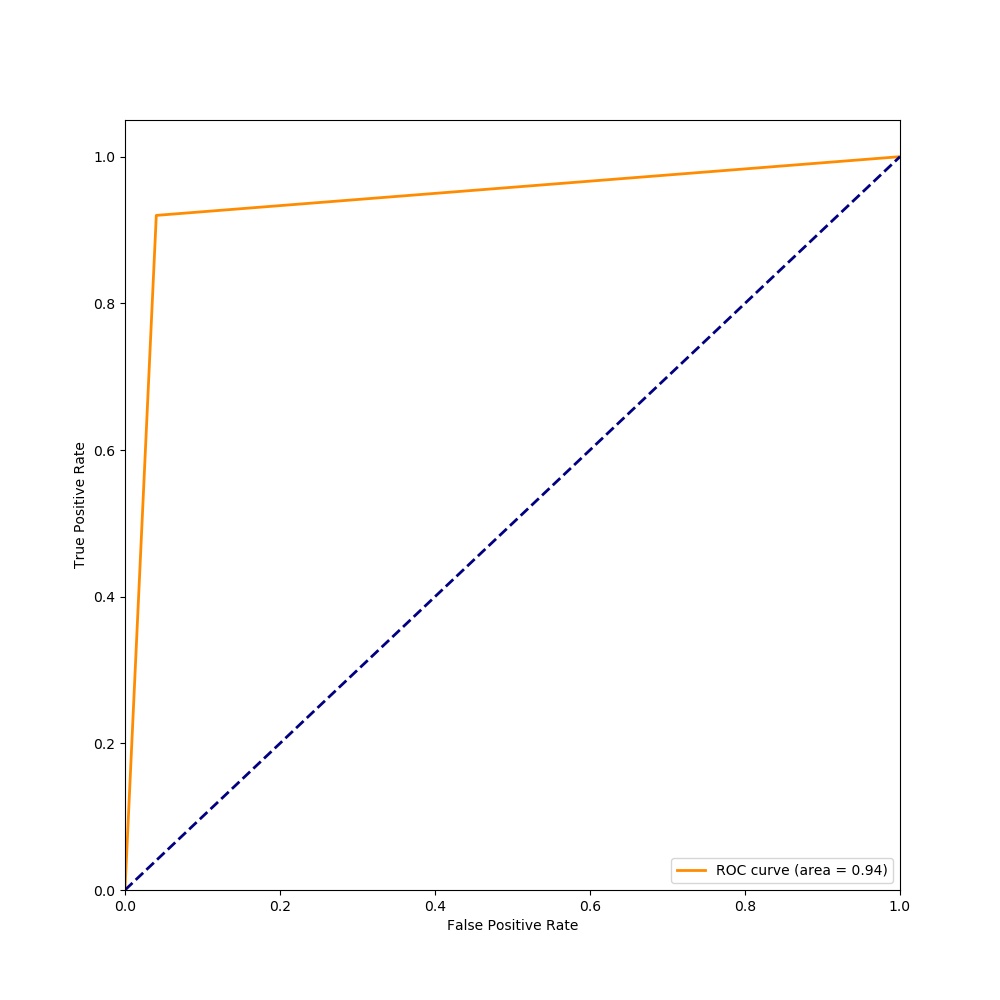}
    \caption{The ROC curve of Decision Tree}
    \end{minipage}
    \end{figure}
    
In terms of accuracy, popular non-Boosting machine learning algorithms, except for random forest, present a relatively low level of performance. Logistic regression, in particular, reaches the lowest level of accuracy at 84.39\%. The performance of MLP classifier is also not ideal, empirically validating the explanation that the credit risk data usually manifests a multi-modal characteristic, which is hard to capture through deep neural networks. On the other hand, random forest as a bagging algorithm, together with boosting-type algorithms, achieves a remarkable level of accuracy, all above 95\% except for lightGBM. Such a high level of performance demonstrates the success of ensemble learning on dataset concerning credit risk, as was mentioned in numerous relevant works.

Within these ensemble algorithms, CatBoost with synthetic features exhibits the best results, reaching up to 95.84\%, while CatBoost without such a combination of features ranks the second. From such a result, we conclude that CatBoost exceeds other present Boosting algorithms in terms of accuracy and that synthetic features can indeed improve the results to an extent. 

The level of AUC follows a similar pattern so we save the explanation here.

\subsection{Feature Importance}
In this part, we measure and visualize the importance of features used in the models. Generally speaking, feature importance measures the contribution a feature makes in the construction of Boosting trees. The more frequently a feature is used in the decision tree, the higher its importance.

When calculating feature importance, we refer to the improvement of classification performance if selecting a specific feature at the $\textit{i}-th$ leaf. The improvement of performance is averaged according to the position of the leaf. The more shallow the leaf is located, i.e. more close to the treetop, the bigger the weight becomes. Eventually, calculate the weighted summation of importance in each tree and average for all Boosting trees, and the feature importance is obtained.

\begin{figure}[htbp]
\centering 
\includegraphics[width = 13.3cm]{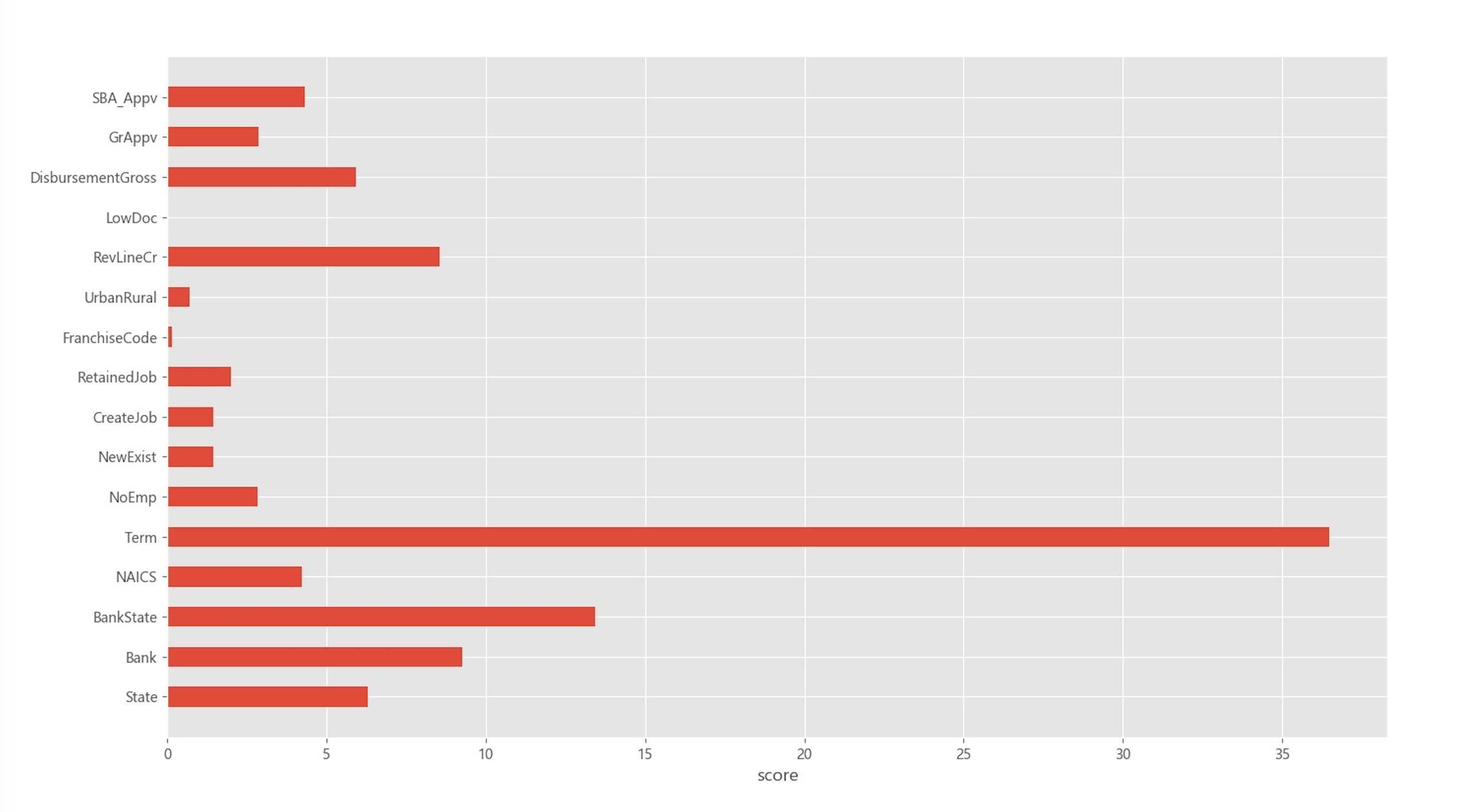}
\caption{Feature importance of CatBoost}
\label{catboost}
\end{figure}

\begin{figure}[htbp]
\centering 
\includegraphics[width = 13.3cm]{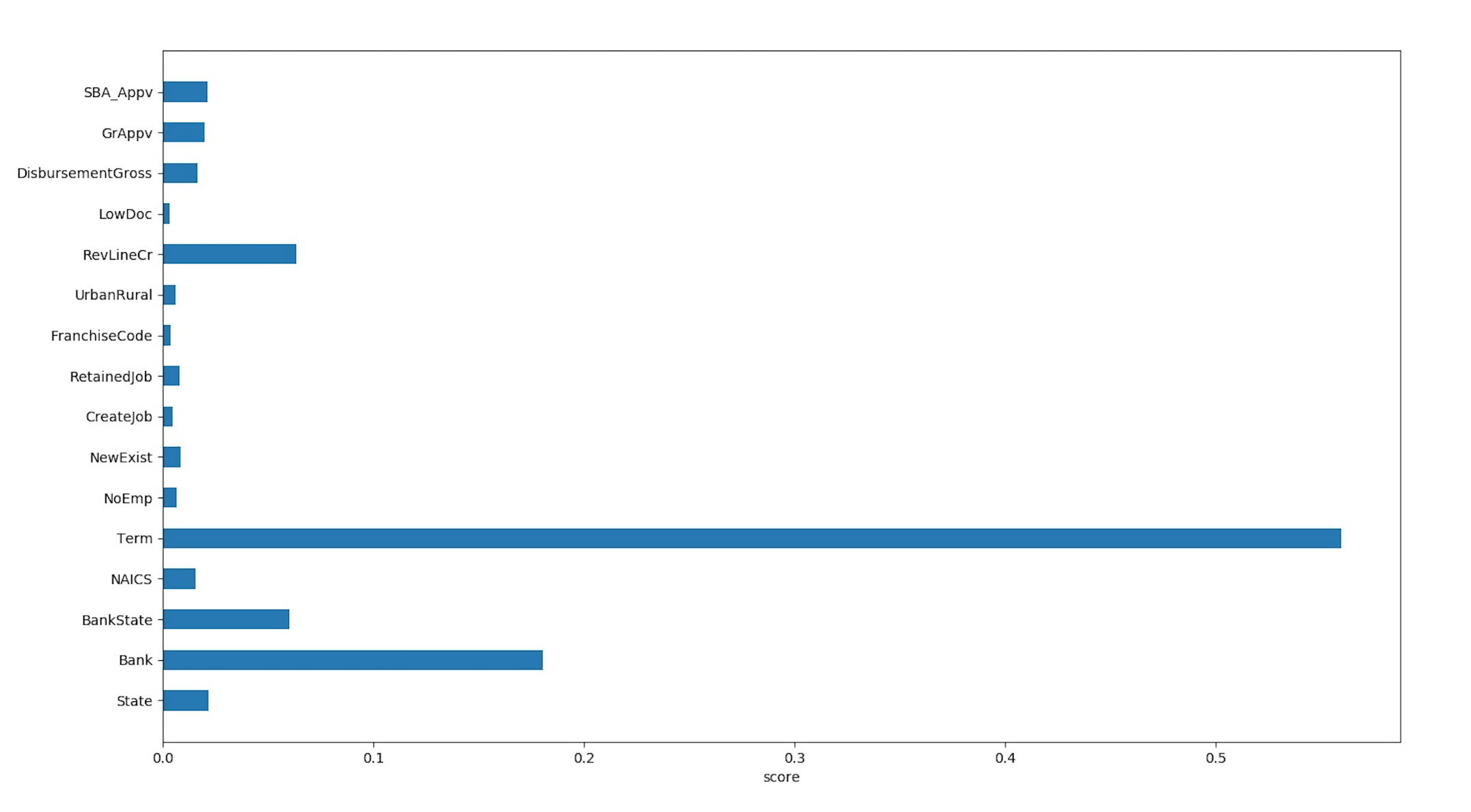}
\caption{Feature importance of XGBoost}
\label{xgboost2}
\end{figure}

Figure \ref{catboost} and \ref{xgboost2} show the feature importance of \textit{national SBA} in XGBoost and CatBoost respectively. It is clear that the feature \textit{term} has the highest feature importance in both models, almost three times the value of the second most important feature. Feature \textit{term} refers to the loan term in months for each loan and the result indicates that the loan term should be deliberately considered by banks and intermediaries. Furthermore, the average values of loan term for default loans and fulfilled loans are computed and are around 45 and 87 months respectively. This can be simply put that the longer the loan term, the lower the default rate on average. 

Bank name and State of bank also appear to be important indicators, although the State of bank has more importance score in CatBoost model and vice versa in XGBoost. Our previous analysis of variables, including the visualization of the distribution of default rate in each State, validates the result to some degree. It is also noteworthy that the gross amount of disbursement, the state of the borrower and whether to revolve line of credit all account for certain part of the importance in CatBoost and should not be left out. On the other hand, for XGBoost, the rest of the factors, except for \textit{RevLineCr}, all contribute to a relatively small part of feature importance.

From a practical perspective, therefore, banks and corresponding institutions should lay importance on the information including the loan term, the State of the borrower, the condition of credit line and carefully grant the amount of disbursement. If, based on the previous records of default rates, the bank suffers from higher average rate of default compared with other local banks or banks from other States, the bank then is suggested to raise the identifying standard and classify borrowers by more strict rules.

\bibliographystyle{plainnat}  
\bibliography{references} 

\begin{thebibliography}{32}
\providecommand{\natexlab}[1]{#1}
\providecommand{\url}[1]{\texttt{#1}}
\expandafter\ifx\csname urlstyle\endcsname\relax
  \providecommand{\doi}[1]{doi: #1}\else
  \providecommand{\doi}{doi: \begingroup \urlstyle{rm}\Url}\fi

\bibitem[A and B(2010)]{2010A}
Ming Yuan Leon~Li A and Peter~Miu B.
\newblock A hybrid bankruptcy prediction model with dynamic loadings on
  accounting-ratio-based and market-based information: A binary quantile
  regression approach - sciencedirect.
\newblock \emph{Journal of Empirical Finance}, 17\penalty0 (4):\penalty0
  818--833, 2010.

\bibitem[Ahad et~al.(2002)Ahad, Fayyaz, and Mehmood]{Ahad}
A.~Ahad, A.~Fayyaz, and T.~Mehmood.
\newblock Speech recognition using multilayer perceptron.
\newblock In \emph{IEEE Students Conference, ISCON '02. Proceedings.},
  volume~1, pages 103--109 vol.1, 2002.
\newblock \doi{10.1109/ISCON.2002.1215948}.

\bibitem[Alfaro-Cortés et~al.(2008)Alfaro-Cortés, Rubio, Gámez, and
  Elizondo]{2008Alfaro}
Esteban Alfaro-Cortés, Noelia Rubio, Matías Gámez, and David Elizondo.
\newblock Bankruptcy forecasting: An empirical comparison of adaboost and
  neural networks.
\newblock \emph{Decision Support Systems}, 45:\penalty0 110--122, 04 2008.
\newblock \doi{10.1016/j.dss.2007.12.002}.

\bibitem[Altman(1968)]{1968FINANCIAL}
E.~I. Altman.
\newblock Financial ratios, discriminant analysis and the prediction of
  corporate bankruptcy.
\newblock \emph{The Journal of Finance}, 23\penalty0 (4):\penalty0 589--609,
  1968.

\bibitem[Beaver(1966)]{1966Financial}
W.~H. Beaver.
\newblock Financial ratios as predictors of failure, empirical reseach in
  accounting: Selected studies.
\newblock \emph{Journal of Accounting Research}, pages 179--199, 1966.

\bibitem[Bellotti et~al.(2019)Bellotti, Brigo, Gambetti, and
  Vrins]{2019ForecastingBellotti}
A.~Bellotti, D.~Brigo, P.~Gambetti, and F.~Vrins.
\newblock Forecasting recovery rates on non-performing loans with machine
  learning.
\newblock In \emph{Credit Scoring and Credit Control XVI}, 2019.

\bibitem[Ben~Jabeur et~al.(2021)Ben~Jabeur, Gharib, Mefteh-Wali, and
  Ben~Arfi]{2021Ben}
Sami Ben~Jabeur, Cheima Gharib, Salma Mefteh-Wali, and Wissal Ben~Arfi.
\newblock Catboost model and artificial intelligence techniques for corporate
  failure prediction.
\newblock \emph{Technological Forecasting and Social Change}, 166:\penalty0
  120658, 05 2021.
\newblock \doi{10.1016/j.techfore.2021.120658}.

\bibitem[Breiman(2001)]{breiman2001random}
Leo Breiman.
\newblock Random forests.
\newblock \emph{Machine learning}, 45\penalty0 (1):\penalty0 5--32, 2001.

\bibitem[Bueyuekkarabacak and Valev(2010)]{2010Bueyuekkarabacak}
Berrak Bueyuekkarabacak and N.~T. Valev.
\newblock The role of household and business credit in banking crises.
\newblock \emph{Journal of Banking \& Finance}, 34\penalty0 (6):\penalty0
  1247--1256, 2010.

\bibitem[Cabrera(1994)]{cabrera1994logistic}
Alberto~F Cabrera.
\newblock Logistic regression analysis in higher education: An applied
  perspective.
\newblock \emph{Higher education: Handbook of theory and research},
  10:\penalty0 225--256, 1994.

\bibitem[Chen et~al.(2016)Chen, Ribeiro, and Chen]{2016Financial}
N.~Chen, B.~Ribeiro, and A.~Chen.
\newblock Financial credit risk assessment: a recent review.
\newblock \emph{Artificial Intelligence Review}, 2016.

\bibitem[Chen and Guestrin(2016)]{chen2016xgboost}
Tianqi Chen and Carlos Guestrin.
\newblock Xgboost: A scalable tree boosting system.
\newblock In \emph{Proceedings of the 22nd acm sigkdd international conference
  on knowledge discovery and data mining}, pages 785--794, 2016.

\bibitem[Coats and Fant(1993)]{1993A}
P.~K. Coats and L.~F. Fant.
\newblock A neural network approach to forecasting financial distress.
\newblock 1993.

\bibitem[Dimitras et~al.(1999)Dimitras, Slowinski, Susmaga, and
  Zopounidis]{1999Business}
A.~I. Dimitras, R.~Slowinski, R.~Susmaga, and C.~Zopounidis.
\newblock Business failure prediction using rough sets.
\newblock \emph{European Journal of Operational Research}, 114\penalty0
  (2):\penalty0 263--280, 1999.

\bibitem[Dorogush et~al.(2018)Dorogush, Ershov, and
  Gulin]{dorogush2018catboost}
Anna~Veronika Dorogush, Vasily Ershov, and Andrey Gulin.
\newblock Catboost: gradient boosting with categorical features support.
\newblock \emph{arXiv preprint arXiv:1810.11363}, 2018.

\bibitem[Hung and Chen(2009)]{2009A}
C.~Hung and J.~H. Chen.
\newblock A selective ensemble based on expected probabilities for bankruptcy
  prediction.
\newblock \emph{Expert Systems with Applications}, 36\penalty0 (3p1):\penalty0
  5297--5303, 2009.

\bibitem[Jones et~al.(2017)Jones, Johnstone, and Wilson]{jones2017predicting}
Stewart Jones, David Johnstone, and Roy Wilson.
\newblock Predicting corporate bankruptcy: An evaluation of alternative
  statistical frameworks.
\newblock \emph{Journal of Business Finance \& Accounting}, 44\penalty0
  (1-2):\penalty0 3--34, 2017.

\bibitem[Ke et~al.(2017)Ke, Meng, Finley, Wang, Chen, Ma, Ye, and
  Liu]{ke2017lightgbm}
Guolin Ke, Qi~Meng, Thomas Finley, Taifeng Wang, Wei Chen, Weidong Ma, Qiwei
  Ye, and Tie-Yan Liu.
\newblock Lightgbm: A highly efficient gradient boosting decision tree.
\newblock \emph{Advances in neural information processing systems},
  30:\penalty0 3146--3154, 2017.

\bibitem[Kumar and Ravi(2006)]{2006Bankruptcy}
P.~R. Kumar and V.~Ravi.
\newblock Bankruptcy prediction in banks by fuzzy rule based classifier.
\newblock In \emph{Digital Information Management, 2006 1st International
  Conference on}, 2006.

\bibitem[Lee(2006)]{2006Genetic}
W.~C. Lee.
\newblock Genetic programming decision tree for bankruptcy prediction.
\newblock In \emph{Joint Conference on Information Sciences}, 2006.

\bibitem[Li et~al.(2018{\natexlab{a}})Li, Mickel, and Taylor]{Li2018}
Min Li, Amy Mickel, and Stanley Taylor.
\newblock “should this loan be approved or denied?”: A large dataset with
  class assignment guidelines.
\newblock \emph{Journal of Statistics Education}, 26\penalty0 (1):\penalty0
  55--66, 2018{\natexlab{a}}.
\newblock \doi{10.1080/10691898.2018.1434342}.
\newblock URL \url{https://doi.org/10.1080/10691898.2018.1434342}.

\bibitem[Li et~al.(2018{\natexlab{b}})Li, Mickel, and Taylor]{table1}
Min Li, Amy Mickel, and Stanley Taylor.
\newblock “should this loan be approved or denied?”: A large dataset with
  class assignment guidelines.
\newblock \emph{Journal of Statistics Education}, 26\penalty0 (1):\penalty0
  55--66, 2018{\natexlab{b}}.
\newblock \doi{10.1080/10691898.2018.1434342}.
\newblock URL \url{https://doi.org/10.1080/10691898.2018.1434342}.

\bibitem[Nanni and Lumini(2006)]{2006An}
L.~Nanni and A.~Lumini.
\newblock An experimental comparison of ensemble of classifiers for biometric
  data.
\newblock \emph{Neurocomputing}, 69\penalty0 (13/15):\penalty0 1670--1673,
  2006.

\bibitem[Prokhorenkova et~al.(2017{\natexlab{a}})Prokhorenkova, Gusev, Vorobev,
  Dorogush, and Gulin]{2017CatBoost}
L.~Prokhorenkova, G.~Gusev, A.~Vorobev, A.~V. Dorogush, and A.~Gulin.
\newblock Catboost: unbiased boosting with categorical features.
\newblock 2017{\natexlab{a}}.

\bibitem[Prokhorenkova et~al.(2017{\natexlab{b}})Prokhorenkova, Gusev, Vorobev,
  Dorogush, and Gulin]{prokhorenkova2017catboost}
Liudmila Prokhorenkova, Gleb Gusev, Aleksandr Vorobev, Anna~Veronika Dorogush,
  and Andrey Gulin.
\newblock Catboost: unbiased boosting with categorical features.
\newblock \emph{arXiv preprint arXiv:1706.09516}, 2017{\natexlab{b}}.

\bibitem[Sigrist and Hirnschall(2019)]{sigrist2019grabit}
Fabio Sigrist and Christoph Hirnschall.
\newblock Grabit: Gradient tree-boosted tobit models for default prediction.
\newblock \emph{Journal of Banking \& Finance}, 102:\penalty0 177--192, 2019.

\bibitem[Son et~al.(2019)Son, Hyun, Phan, and Hwang]{SON2019112816}
H.~Son, C.~Hyun, D.~Phan, and H.J. Hwang.
\newblock Data analytic approach for bankruptcy prediction.
\newblock \emph{Expert Systems with Applications}, 138:\penalty0 112816, 2019.
\newblock ISSN 0957-4174.
\newblock \doi{https://doi.org/10.1016/j.eswa.2019.07.033}.
\newblock URL
  \url{https://www.sciencedirect.com/science/article/pii/S0957417419305123}.

\bibitem[Vapnik et~al.(1995)Vapnik, Guyon, and Hastie]{vapnik1995support}
Vladimir Vapnik, Isabel Guyon, and Trevor Hastie.
\newblock Support vector machines.
\newblock \emph{Mach. Learn}, 20\penalty0 (3):\penalty0 273--297, 1995.

\bibitem[Zhang et~al.(1999)Zhang, Hu, Patuwo, and
  Indro]{Guoqiang1999Artificial}
Guoqiang Zhang, Michael~Y. Hu, B~Eddy Patuwo, and Daniel~C. Indro.
\newblock Artificial neural networks in bankruptcy prediction: General
  framework and cross-validation analysis.
\newblock \emph{European Journal of Operational Research}, 1999.

\bibitem[Zhang et~al.(2013)Zhang, Wang, and Ji]{2013A}
Y.~Zhang, S.~Wang, and G.~Ji.
\newblock A rule-based model for bankruptcy prediction based on an improved
  genetic ant colony algorithm.
\newblock \emph{Mathematical Problems in Engineering,2013,(2013-11-28)},
  2013\penalty0 (pt.14):\penalty0 1--10, 2013.

\bibitem[Zieba et~al.(2016{\natexlab{a}})Zieba, Tomczak, and
  Tomczak]{2016EnsembleZieba}
M.~Zieba, S.~K. Tomczak, and J.~M. Tomczak.
\newblock Ensemble boosted trees with synthetic features generation in
  application to bankruptcy prediction.
\newblock \emph{Expert Systems with Applications}, 58\penalty0 (Oct.):\penalty0
  93--101, 2016{\natexlab{a}}.

\bibitem[Zieba et~al.(2016{\natexlab{b}})Zieba, Tomczak, and
  Tomczak]{2016Zieba}
M.~Zieba, S.~K. Tomczak, and J.~M. Tomczak.
\newblock Ensemble boosted trees with synthetic features generation in
  application to bankruptcy prediction.
\newblock \emph{Expert Systems with Applications}, 58\penalty0 (Oct.):\penalty0
  93--101, 2016{\natexlab{b}}.

\end{thebibliography}

\begin{appendices}
\section{Tuning hyperparameters for models}

\begin{table}[H]
\label{tuning results}
\centering
\begin{tabular}{@{}lll@{}}
\toprule
\textit{\textbf{Model}}          & \textit{\textbf{Optimal parameters}}                                                                                                                                                                                                                                          & \textit{\textbf{Model result}}                                           \\ \midrule
\textbf{CatBoost with synthetic features} & \begin{tabular}[c]{@{}l@{}}Depth: 10 \\ loss\_function: Logloss \\ learning\_rate: 0.05\end{tabular}                                                                                                                                                                          & \begin{tabular}[c]{@{}l@{}}Accuracy: 95.74\%\\ AUC: 98.59\%\end{tabular} \\
\textbf{}                        &                                                                                                                                                                                                                                                                               &                                                                          \\
\textbf{CatBoost}                & \begin{tabular}[c]{@{}l@{}}Depth: 10 \\ loss\_function: Logloss  \\ learning\_rate: 0.05\end{tabular}                                                                                                                                                                         & \begin{tabular}[c]{@{}l@{}}Accuracy: 95.74\%\\ AUC: 98.59\%\end{tabular} \\
\textbf{}                        &                                                                                                                                                                                                                                                                               &                                                                          \\
\textbf{XGBoost}                 & \begin{tabular}[c]{@{}l@{}}N\_estimators:62 \\ max\_depth:9 98.59\% \\ Min\_child\_weight: 2 \\ Gamma: 0 \\ colsample\_bytree: 0.7 \\ subsample 0.9 \\ reg\_alpha: 3 \\ reg\_lambda: 3\end{tabular}                                                                           & \begin{tabular}[c]{@{}l@{}}Accuracy: 95.56\%\\ AUC: 98.53\%\end{tabular} \\
\textbf{}                        &                                                                                                                                                                                                                                                                               &                                                                          \\
\textbf{lightGBM}                & \begin{tabular}[c]{@{}l@{}}N\_estimators: 127 \\ max\_depth: 6 \\ num\_leaves: 10 \\ max\_bin: 95\\ min\_data\_in\_leaf: 81 \\ bagging\_fraction: 0.6 \\ bagging\_freq: 0\\ feature\_fraction: 1.0 \\ lambda\_l1: 0.7\\ lambda\_l2: 0.3 \\ min\_split\_gain: 0.0\end{tabular} & \begin{tabular}[c]{@{}l@{}}Accuracy: 94.74\%\\ AUC: 98.15\%\end{tabular} \\
\textbf{}                        &                                                                                                                                                                                                                                                                               &                                                                          \\
\textbf{MLPclassifier}           & \begin{tabular}[c]{@{}l@{}}Activation: logistic \\ Solver: adam\\ Hidden\_layer\_sizes: (10,10)\end{tabular}                                                                                                                                                                  & \begin{tabular}[c]{@{}l@{}}Accuracy: 92.91\%\\ AUC: 97.26\%\end{tabular} \\
\textbf{}                        &                                                                                                                                                                                                                                                                               &                                                                          \\
\textbf{Random Forest}           & \begin{tabular}[c]{@{}l@{}}N\_estimators: 170\\ Max\_depth: 16\\ Max\_features: 8\end{tabular}                                                                                                                                                                                & \begin{tabular}[c]{@{}l@{}}Accuracy: 95.53\%\\ AUC: 98.52\%\end{tabular} \\
\textbf{}                        &                                                                                                                                                                                                                                                                               &                                                                          \\
\textbf{SVM}                     & \begin{tabular}[c]{@{}l@{}}kernel: rbf \\ C:7 \\ Gamma: 0.1\end{tabular}                                                                                                                                                                                                      & \begin{tabular}[c]{@{}l@{}}Accuracy: 91.42\%\\ AUC: 96.21\%\end{tabular} \\
\textbf{}                        &                                                                                                                                                                                                                                                                               &                                                                          \\
\textbf{Logistic Regression}     & \begin{tabular}[c]{@{}l@{}}penalty: I1 \\ solver: liblinear\end{tabular}                                                                                                                                                                                                      & \begin{tabular}[c]{@{}l@{}}Accuracy: 84.39\%\\ AUC: 90.09\%\end{tabular} \\ \bottomrule
\end{tabular}
\end{table}

\section{Reproducible experiments}
Reproducible codes for this experiment can be downloaded at

\begin{center}
    \href{https://github.com/wanghaoxue0/Catboost-synthetic}{https://github.com/wanghaoxue0/Catboost-synthetic}

\end{center}

\end{appendices}

\end{document}